\documentclass[fleqn,usenatbib]{mnras}

\usepackage{newtxtext,newtxmath}

\usepackage[T1]{fontenc}

\DeclareRobustCommand{\VAN}[3]{#2}
\let\VANthebibliography\thebibliography
\def\thebibliography{\DeclareRobustCommand{\VAN}[3]{##3}\VANthebibliography}

\usepackage{graphicx}	
\usepackage{amsmath}	
\usepackage{caption}
\usepackage{subcaption}
\usepackage{multirow}
\usepackage{relsize}

\title[No galaxy-wide outflows in F13451+1232]{No evidence for fast, galaxy-wide ionised outflows in a nearby quasar --- the importance of accounting for beam smearing}

\author[L. R. Holden \& C. N. Tadhunter]
{Luke R. Holden$^{1}$,\thanks{E-mail: l.holden@herts.ac.uk}
Clive N. Tadhunter$^{2}$
\\
$^{1}$Centre for Astrophysics Research, University of Hertfordshire, Hatfield, AL10 9AB, UK. \\
$^{2}$Department of Physics $\&$ Astronomy, University of Sheffield, S6 3TG Sheffield, UK. \\
}

\date{Accepted XXX. Received YYY; in original form ZZZ}

\pubyear{2023}

\begin{document}
\label{firstpage}
\pagerange{\pageref{firstpage}--\pageref{lastpage}}
\maketitle

\begin{abstract}
	To test the scenario that outflows accelerated by active galactic nuclei (AGN) have a major impact on galaxy-wide scales, we have analysed deep VLT/MUSE data for the type-2 quasar/ultraluminous infrared galaxy F13451+1232 --- an object that represents the major mergers considered in models of galaxy evolution. After carefully accounting for the effects of atmospheric seeing that had smeared the emission from known compact nuclear outflows across the MUSE field of view, we find that the large-scale kinematics in F13451+1232 are consistent with gravitational motions that are expected in a galaxy merger. Therefore, the fast ($\mathrm{W_{80}}>500$\;km\;s$^{-1}$) warm-ionised AGN-driven outflows in this object are limited to the central $\sim$100\;pc of the galaxy, although we cannot rule out larger-scale, lower-velocity outflows. Moreover, we directly demonstrate that failure to account for the beam-smearing effects of atmospheric seeing would have led to the mass outflow rates and kinetic powers of spatially-extended emission being overestimated by orders of magnitude. We also show that beam-smeared compact-outflow emission can be significant beyond radial distances of 3.5\;arcseconds (more than eight times the radius of the seeing disk), and support the argument that some previous claims of large-scale outflows in active galaxies were likely the result of this effect rather than genuine galaxy-wide ($r>5$\;kpc) outflows. Our study therefore provides further evidence that warm-ionised AGN-driven outflows are limited to the central kiloparsecs of galaxies and highlights the critical importance of accounting for atmospheric seeing in ground-based observational studies of active galaxies.
\end{abstract}

\begin{keywords}
galaxies: active -- galaxies: evolution -- galaxies: individual: F13451+1232 --- ISM: jets and outflows --- quasars: general --- galaxies: interactions
\end{keywords}



\section{Introduction}
\label{section: introduction}
Models of galaxy evolution now routinely require active galactic nucleus (AGN) feedback --- the heating and ejection of gas in host galaxies by radiation, jets, and winds --- in order to reproduce the observed properties of the local galaxy population (e.g. \citealt{Schaye2015, Dave2019, Zinger2020}) and empirical scaling relations between supermassive black holes (SMBHs) and host-galaxy bulges (e.g. \citealt{Silk1998, King2003, DiMatteo2005, Hopkins2010}). In such models, outflows of gas that are accelerated by AGN often extend to galaxy-wide scales ($r>5$\;kpc: e.g. \citealt{Silk1998, DiMatteo2005, Curtis2016, Barai2018, Costa2018, Costa2022, Zubovas2023}), therefore having a direct impact on the global star formation efficiency of the host galaxy that is not limited to the near-nuclear regions.

In agreement with this scenario, some studies that make use of ground-based integral field unit (IFU) spectroscopy of nearby quasars have claimed evidence for outflows in the warm ionised gas phase ($10,000<T<25,00$\;K; traced by emission lines such as [O III]$\lambda\lambda4959,5007$ and H$\beta$) that extend up to radial distances of tens of kiloparsecs (e.g. \citealt{Fu2009, Westmoquette2012, Liu2013, Liu2014, Harrison2014, McElroy2015, Wylezalek2017}). In contrast, other studies of nearby active galaxies (galaxies hosting AGN) that make use of ground-based long-slit spectroscopy (e.g. \citealt{Das2006, VillarMartin2016, Spence2018, Rose2018, Santoro2020}) and space-based imaging (e.g. \citealt{Tadhunter2018}) and spectroscopy (e.g. \citealt{Fischer2018, Tadhunter2019}) have instead found warm-ionised outflows to extend to maximum radial distances of a few kiloparsecs from the central AGN. 

A possible reason for this apparent discrepancy is that ground-based IFU observations suffer from the beam-smearing effects of atmospheric seeing, which may artificially spread emission from compact, spatially-unresolved nuclear outflows across the field of the view (FOV) of the observations. This effect was investigated by \citet{Husemann2016}, who found that atmospheric seeing may lead to overestimations of extended-narrow-line-region (ENLR: $r>1$\;kpc) radii by up to a factor of two (see also \citealt{Hainline2014}). However, since it is not clear whether ENLR emission represents outflowing gas or AGN-photoionised gas undergoing (non-outflowing) gravitational motions, the impact of atmospheric seeing on direct measurements of \textit{outflow} radii may be greater.

On the other hand, some studies which claim that AGN-driven outflows are relatively compact make use of techniques such as spectroastromentry (e.g. \citealt{VillarMartin2016, Santoro2018, Santoro2020}) and Hubble Space Telescope (HST) imaging and long-slit spectroscopy (e.g. \citealt{Fischer2018, Tadhunter2018, Tadhunter2019}), which are sensitive to high-surface-brightness emission in near-nuclear regions but potentially insensitive to larger-scale, lower-surface-brightness emission. Indeed, as argued by \citet{Spence2018}, it is possible for a spatially-extended, off-nuclear outflow to have a much lower surface brightness --- yet carry significantly more mass (provided that its density is sufficiently low) --- than a compact, nuclear outflow.

In addition to galaxy-wide outflows, models of galaxy evolution typically involve a fraction of the radiation produced by AGN coupling to the interstellar medium (typically 0.5--5\;per\;cent: e.g. \citealt{DiMatteo2005, Hopkins2010, Schaye2015, Dubois2016})\footnote{Due to complexities regarding gas cooling, properties of the host ISM, and gravitational effects, not all of the AGN radiation that is coupled to the ISM in these models would be expected to become outflow kinetic power (see discussions in \citealt{Harrison2018} and \citealt{Harrison2024}). Furthermore, it is crucial to note that other gas phases may contain most of the outflowing gas mass and kinetic power (see \citealt{Cicone2018}), as is the case for the nuclear outflows in F13451+1232.}. However, detailed observational studies of nearby active galaxies that utilised robust diagnostics of key warm-ionised outflow properties have often derived kinetic powers that are far below this fraction (e.g. \citealt{Holt2011, Rose2018, Santoro2018, Santoro2020, Baron2019b, Revalski2021, Holden2023, HoldenTadhunter2023, Speranza2024, Bessiere2024}). Crucially, these studies relied on methods that may not have been sensitive to low-density, spatially-extended outflow emission. As discussed earlier, such a component may carry a significantly higher mass (and hence kinetic power) than the dense gas traced in those studies, and therefore could change the interpretation of the impact of outflows on their host galaxies. Clearly, the existence (or lack thereof) of spatially-extended ($r>5$\;kpc), low-surface-brightness, tenuous outflows in the warm-ionised phase now needs to be directly verified.

Ultraluminous infrared galaxies (ULIRGs; $L_{5-500\;\mu\mathrm{m}}>10^{12}$\;L$_\odot$: \citealt{Sanders1996}) are ideal objects to search for the putative large-scale, low-surface-brightness outflows: they represent the peaks of galaxy mergers and therefore, according to models of galaxy evolution that invoke AGN triggered in such conditions, are expected to host prominent, galaxy-wide outflows \citep{DiMatteo2005, Hopkins2008, Johansson2009}. In particular, the ULIRG F13451+1232 (also known as 4C12.50) is an excellent laboratory for this type of study. Not only does it have one of the most well-characterised multi-phase AGN-driven outflows, which is detected in compact ($r<100$\;pc) cold-molecular \citep{Holden2024}, neutral-atomic \citep{Morganti2013_4c1250}, and warm-ionised \citep{Holt2003, Holt2011, Rose2018, VillarMartin2023} emission near the primary nucleus, but it is classified as type-2 quasar (QSO2; $L_\mathrm{bol}=4.8\times10^{45}$\;erg\;s$^{-1}$: \citealt{Rose2018}) and also hosts a luminous ($L_\mathrm{1.4\;GHz}=1.9\times10^{26}$\;W\;Hz$^{-1}$), compact ($r\sim130$\;pc) radio source \citep{Stanghellini1997, Lister2003, Morganti2013_4c1250}. Hence, both of the main AGN-driven-outflow acceleration mechanisms --- a strong radiation field, and powerful jets --- are present. In this context, it is perhaps surprising that the mass outflow rates and kinetic powers that are derived for the near-nuclear warm-ionised outflows in F13451+1232 ($\dot{M}_\mathrm{out}=3.0$--11.3\;M$_\odot$; $\dot{E}_\mathrm{kin}=(0.2$--$2.4)\times10^{43}$\;erg\;s$^{-1}=0.04$--0.49\;per\;cent of $L_\mathrm{bol}$: \citealt{Rose2018}) are relatively modest compared to the coupling factors used in many theoretical models (e.g. \citealt{DiMatteo2005, Hopkins2010, Schaye2015}). However, following the arguments made by \citet{Spence2018}, it is possible that a galaxy-wide, warm-ionised outflow component exists that has much lower densities while being significantly more massive, and has remained undetected in the nuclear regions due to emission from the high-density gas dominating on these scales.

Crucially, HST imaging observations demonstrate that the near-nuclear warm ionised outflow in PKS1345+12 is compact ($r_\mathrm{[OIII]}\sim69$\;pc: \citealt{Tadhunter2018}) and unresolved in ground-based observations. Coupled with its high flux and extreme kinematics, this allows the point spread function (PSF) of the nuclear outflow to be accurately determined and subtracted from the field of view, thereby facilitating searches for any extended, low-surface-brightness outflow emission.

Therefore, to directly test if a galaxy-wide component to the warm ionised outflow phase exists in a ULIRG/QSO2 --- and to determine the impact of atmospheric seeing on AGN-driven-outflow spatial extents, kinematics, masses, and kinetic powers --- here, we analyse deep, wide-field Very Large Telescope (VLT) / Multi Unit Spectroscopic Explorer (MUSE) observations of F13451+1232.

This study is presented as follows. In Section \ref{section: observations_and_data_reduction}, we describe the dataset and the data reduction process, in addition to atmospheric seeing measurements for the observations. Our methodology and the analysis of the data is presented in Section \ref{section: analysis_and_results}, and in Section \ref{section: discussion} we discuss the interpretation of these results and their implications for previous and future work regarding AGN-driven outflows. Finally, we give our conclusions in Section \ref{section: conclusions}.

We assume a cosmology of $H_0=70$\;km\;s$^{-1}$\;Mpc$^{-1}$, $\Omega_\mathrm{m}=0.3$, and $\Omega_\mathrm{\lambda}=0.7$ throughout this work. At the redshift of F13451+1232 ($z=0.121680$: \citealt{Lamperti2022}), this corresponds\footnote{Calculated using Ned Wright's Javascript Cosmology Calculator \citep{Wright2006}.} to an arcsecond-to-kpc spatial conversion factor of 2.189\;kpc/arcsec and a luminosity distance of $D_\mathrm{L}=570$\;Mpc.

\section{Observations and data reduction}
\label{section: observations_and_data_reduction}

\subsection{Archival VLT/MUSE-DEEP data}
\label{section: observations_and_data_reduction: observations}

Archival MUSE-DEEP\footnote{\url{https://doi.eso.org/10.18727/archive/42}} data products of F13451+1232 were downloaded from the ESO Archive Science Portal\footnote{\url{https://archive.eso.org/scienceportal/home}}. The MUSE-DEEP data project combines observations from multiple observing blocks (not necessarily taken contiguously, or on the same night) by first removing instrumental signatures (i.e. bias subtraction, dark subtraction, flat-fielding, telluric correction, and sky subtraction) from each constituent science exposure and performing astrometry, flux, and wavelength calibration using the \textsc{MUSE Data Reduction Pipeline}\footnote{Version 2.8 of the MUSE Data Reduction Pipeline was used to reduce the MUSE-DEEP dataset for F13451+1232.} \citep{Weilbacher2020}. The reduced exposures are then spatially aligned, and overlapping pixels are resampled to produce a deep datacube.

\begin{figure}
    \centering
    \includegraphics[width=1\linewidth]{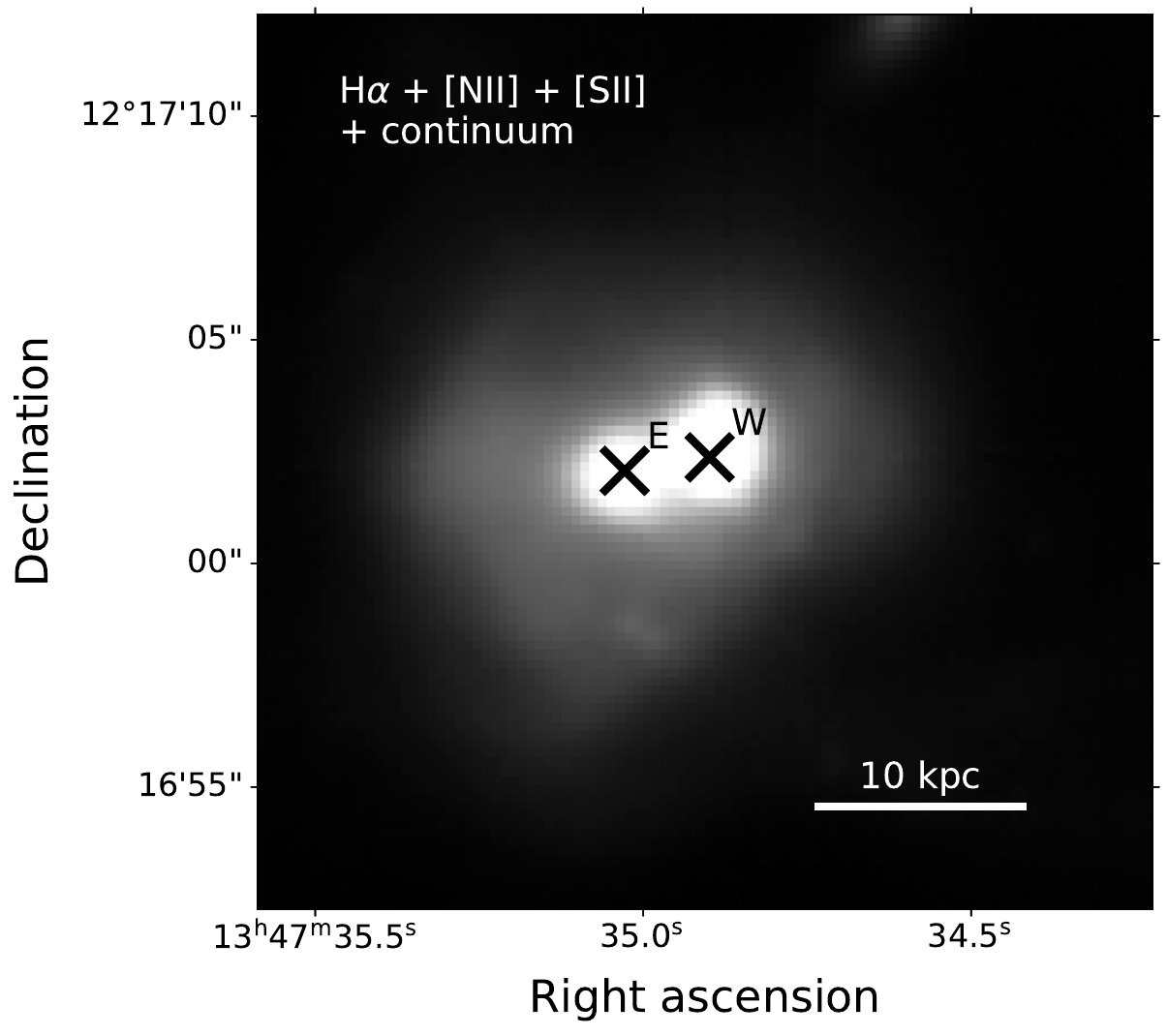}
    \caption{6400--6850\;{\AA} rest-wavelength image (covering H$\alpha$ + [NII]$\lambda\lambda$6548,6584, [SII]$\lambda\lambda$6717,6731, and continuum) of F13451+1232, produced from the archival MUSE-DEEP data. The primary (western: "W") and secondary (eastern: "E") nuclei are marked with black crosses.}
    \label{fig: observations_and_data_reduction: halpha_sii_image}
\end{figure}

The observations used to produce the MUSE-DEEP cube for F13451+1232 were taken as part of ESO programme 0103.B-0391 (PI Arribas) on the 11th and 12th May 2019 using the instrument's Wide Field Mode (WFM) with adaptive optics (AO), centred on the object's bright primary (western) nucleus (Figure \ref{fig: observations_and_data_reduction: halpha_sii_image}). MUSE-WFM covers a field of view of $1.53\times1.53$\;arcminutes with a pixel scale of 0.2\;arcseconds per pixel, and has a wavelength range of 4700--9350\;{\AA} with a spectral resolution of $\sim$1800 at 5000\;{\AA}. The total combined exposure time of the deep datacube was 6732\;seconds.

\subsection{Further reduction of MUSE-DEEP data}
\label{section: observations_and_data_reduction: further_reduction}

To remove residuals of the \textsc{MUSE Data Reduction Pipeline} sky-subtraction routine, a second-order sky subtraction was performed on the deep datacube. First, we selected a $0.6\times0.6$\;arcsecond region at a radial distance of $\sim$20\;arcseconds northwest of the primary nucleus that was free of emission, and took the median of the spaxels in this region to give a median spectrum. Since the data had already been (first-order) sky-subtracted by the pipeline, the only features in this spectrum were residuals of subtracted telluric lines. Second, this median spectrum was subtracted from each spaxel in the deep datacube, resulting in a second-order-sky-subtracted cube.

Finally, each spaxel was corrected for Galactic extinction by using the $R_\mathrm{v}=3.1$ extinction law from \citet{Cardelli1989} with the mean value of $\mathrm{E(B}-\mathrm{V)}_\mathrm{mean}=0.0286\pm0.0005$ that was found in the direction of F13451+1232 (taken from the extinction maps produced by \citealt{Schlegel1998} and recalibrated by \citealt{Schlafly2011}).

\subsection{Atmospheric seeing estimates}
\label{section: observations_and_data_reduction: seeing}

To quantify the atmospheric seeing of the dataset, we extracted a $6\times6$\;arcsecond region from the deep cube around a star that lies in the field of view at a radial distance of $\sim$17\;arcseconds in projection from the primary nucleus of F13451+1232. The flux density of this star was integrated between the wavelengths 5496--5664\;{\AA} (corresponding to the [OIII]$\lambda\lambda$4595,5007 doublet) to produce a continuum image. Fitting a two-dimensional (2D) Moffat profile --- which accurately describes a seeing disk \citep{Moffat1969} and the MUSE-WFM AO-reduced point spread function \citep{Fusco2020} --- to this image gave a corresponding full width at half maximum of FWHM$_{\star,\mathrm{[OIII]}}=0.79\pm0.10$\;arcseconds, which is taken to be the AO-reduced seeing value for the deep datacube at 5000\;{\AA}. The seeing value derived in this way is consistent with the values measured by the VLT observatory DIMM during the constituent observations (0.66--0.91\;arcseconds)\footnote{VLT observatory DIMM seeing values queried using the DIMM Seeing Query Form: \url{http://archive.eso.org/wdb/wdb/asm/dimm_paranal/form}}.

\section{Analysis and results}
\label{section: analysis_and_results}

\subsection{A Bayesian emission-line fitting routine}
\label{section: analysis_and_results: bayesian_emission_line_fitting_routine}

In order to fit the large number of emission lines involved in this analysis robustly, we have developed an automated, Bayesian emission-line fitting routine. This routine uses the \textsc{emcee} \textsc{Python} module \citep{FormanMackey2013} --- which is an implementation of the Affine Invariant Markov Chain Monte Carlo (MCMC) Ensemble sampler \citep{Goodman2010} --- to fit a first-order polynomial to the continuum and $N_\mathrm{g}$ Gaussian components to a given emission-line profile. In this routine, $N_\mathrm{g}$ is iteratively increased from zero, and at each iteration, the posterior odds ratio of the current and previous iteration is used to determine if another iteration (one more Gaussian component) statistically improves the fit to the line profile, thus preventing over-fitting.

Lines, doublets, and emission-line blends (such as H$\alpha$ + {[}NII{]}$\lambda\lambda$6548,6584) were fit individually; a sufficient wavelength range of continuum on either side of the line profile(s) was included in the fits. In cases where the modelled lines arise from a doublet of the same ion (e.g. [SII]$\lambda\lambda$6717,6731), the width of a given Gaussian component was set to be the same for each doublet line, and the wavelength separations were set to those defined by atomic physics \citep{Osterbrock2006}. Furthermore, for lines arising from the same upper energy level, the flux ratios were also set to those defined by atomic physics ([OIII]$(5007/4959)=2.99$; [NII]$(6584/6548)=2.92$; as determined with the \textsc{PyNeb Python} module: \citealt{Luridiana2015}).

Before the main production run of the MCMC routine, a burn-in phase was performed. The steps of this phase were not included in determining the final parameters of the fit, as their purpose was to ensure that the walkers started in a region of probability that is more representative of the sampled parameter distribution. 1000 steps were used for both the burn-in and the main production run, as it was found that the walkers converged well before this number of steps; 200 walkers were used in both cases.

The initial starting points for the model parameters were set by Gaussian distributions around physically-motivated estimates: the continuum flux offset (i.e. the $y$-intercept of the first-order polynomial) was set to be the mean flux measured on either side of the emission line; the continuum slope (the gradient of the first-order polynomial) was set to zero; the peak value of each Gaussian component was set as half of the maximum flux value seen in the data, the Gaussian centroids were set to the expected wavelength of the emission line in the rest frame of the galaxy (defined by the object's redshift measured using CO(2--1) observations by \citealt{Lamperti2022}), and the Gaussian widths were set to be twice the instrumental line width of MUSE ($\sigma_\mathrm{inst}=1.1$\;{\AA}: \citealt{Weilbacher2020}).

Priors for the routine were also physically motivated. For the Gaussian components, their peak values were required to be equal-to or greater-than zero (to ensure only emission was being modelled), their centroids were not allowed to be more than 50\;{\AA} in separation from the rest wavelength of the line in the galaxy's rest frame (corresponding to ${v}\sim3000$\;km\;s$^{-1}$ at 5000\;{\AA}), and their widths were constrained to be greater than the MUSE instrumental width. Furthermore, the flux ratios for lines arising from the same ions (but with different upper energy levels) were required to be within the ratio limits defined by atomic physics --- for example, 0.44\;\textless\;[SII](6717/6731)\;\textless\;1.45 (determined using the \textsc{PyNeb Python} module). 

The log-likelihood function used in the MCMC fits was
\begin{equation}
    \ln L=\frac{1}{2}\mathlarger{\sum^k_{i=1}}\Biggl(\frac{F_\mathrm{\lambda,i}-F^\mathrm{m}_\mathrm{\lambda,i}}{F^\mathrm{err}_\mathrm{\lambda,i}}\Biggl)^2,
    \label{eq: analysis_and_results: chisq}
\end{equation}
where, at the wavelength step $i$ (and up to the final wavelength step $k$), $F_\mathrm{\lambda,i}$ is the observed flux density, $F^\mathrm{m}_\mathrm{\lambda,i}$ is the modelled flux density, and ${F^\mathrm{err}_\mathrm{\lambda,i}}$ is the uncertainty associated with the observed flux density.

The fitting routine begins with $N_\mathrm{g}=0$ (i.e. only a first-order polynomial), for which a fit to the data is produced using the MCMC ensemble. The posterior probabilities for the initial run are recorded, and this process is repeated for $N_\mathrm{g}=N_\mathrm{g}+1$ until the ratio of posterior probabilities (Bayesian odds) of successive runs is less than two\footnote{As a check on this criterion, we also performed this routine with the Bayesian Information Criterion (BIC), where the more-complex model was chosen if the difference of BIC values between successive runs was greater than 6. When applied to the MUSE-DEEP data, the resulting model parameters were not significantly different from those produced using the Bayesian odds criterion.}; if this condition is not met, the routine continues until it is fulfilled. In this way, the minimum number of statistically-meaningful Gaussian components required to describe the emission-line profiles is determined by the routine, avoiding over-fitting. 

Cases where only a first-order polynomial was required to adequately describe the spectrum (i.e. $N_\mathrm{g}=0$) were considered to be non-detections, whereas in cases that required one or more Gaussian components, the value for each model parameter was taken to be the 50th percentile for the marginalised probability distribution (i.e. the probability distribution for each parameter), and the $1\sigma$ uncertainties were taken as the 16th and 84th percentiles.

\subsection{The effect of atmospheric seeing on outflow extents and kinematics}
\label{section: analysis_and_results: seeing}

\subsubsection{Nuclear aperture extraction and modelling}
\label{section: analysis_and_results: seeing: nuclear_aperture_extraction}

\begin{figure}
    \centering
    \includegraphics[width=\linewidth]{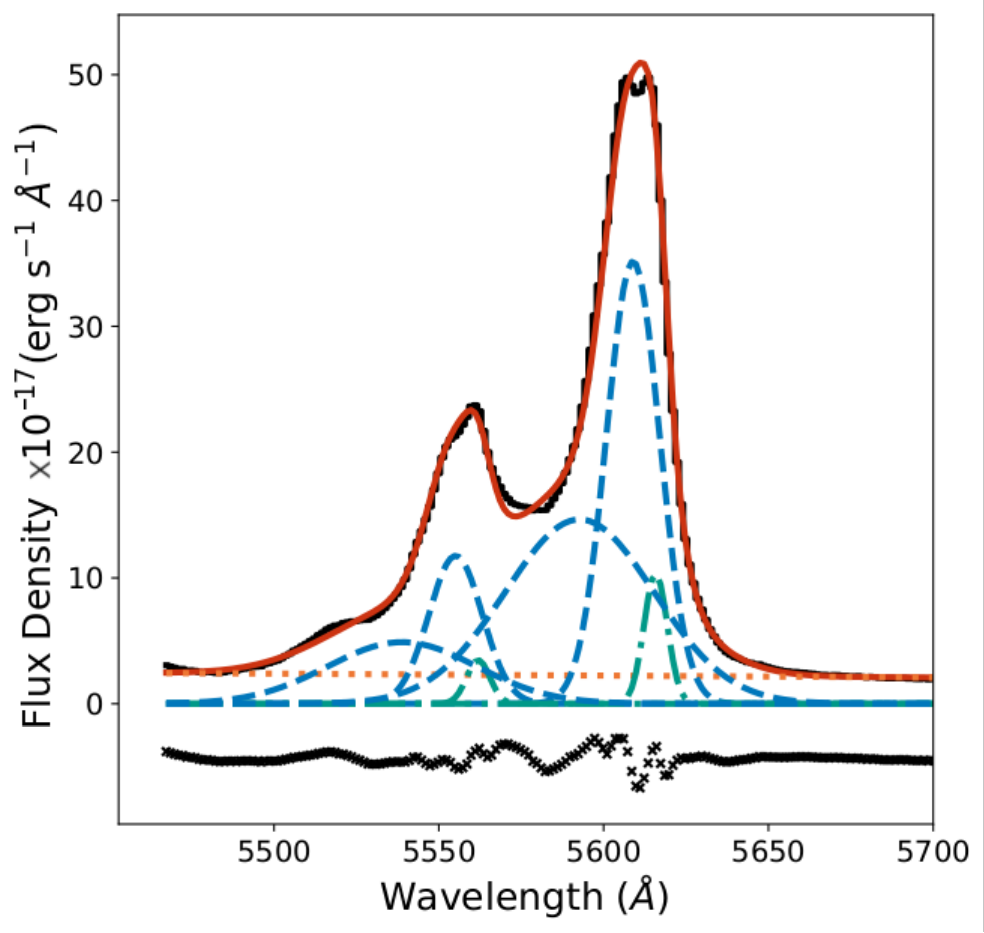}
    \caption{[OIII]$\lambda\lambda$4959,5007 line profile for the extracted circular $r=0.4$\;arcsec aperture centred on the primary nucleus of F13451+1232 (black solid line). The overall fit to the line profile is shown as a solid red line; the broad Gaussian components of this fit (the `nuclear model') are shown as dashed blue lines, the narrow Gaussian component is shown as a dash-dotted green line, and the first-order polynomial (accounting for the continuum) is shown as a dotted orange line. Residuals (flux $-$ model) are shown as crosses below the line profile.}
    \label{fig: analysis_and_results: seeing: nuclear_aperture_spectrum}
\end{figure}

Given that compact ($r$\;\textless\;100\;pc), luminous, warm-ionised outflows have been detected and characterised near the primary nucleus of F13451+1232 \citep{Holt2003, Holt2011, Rose2018, Tadhunter2018}, it is possible that the beam-smearing effects of atmospheric seeing had artificially spread this emission to larger spatial scales in the MUSE-DEEP datacube FOV. To account for this, we first extracted a circular aperture centred on the primary nucleus, the radius of which was set to be 0.4\;arcseconds ($\sim$0.9\;kpc), corresponding to the half width at half maximum ($\mathrm{HWHM}=\mathrm{FWHM}/2$) of the seeing value for the dataset (HWHM$_{\star,\mathrm{[OIII]}}=0.40\pm0.10$ arcseconds: Section\;\ref{section: observations_and_data_reduction: seeing}). This radius was selected because intermediate and broad components of any line profiles within it are expected to be due to the prominent compact outflows ($r_\mathrm{[OIII]}\sim69$\;pc: \citealt{Tadhunter2018}), and because there was sufficient signal for the robust modelling and fitting of these profiles. The spaxels contained in the nuclear aperture were summed to give a total nuclear spectrum, while the flux uncertainties of each spaxel were added in quadrature. Then, the Bayesian emission-line-fitting routine was used to produce and fit a model to the [OIII]$\lambda\lambda4959,5007$ doublet of the nuclear spectrum. To ensure that the model accounted for the extended blue wings and dual-peaked line profiles seen in the nuclear spectrum, the results of a least-squares fit consisting of three Gaussian components were used as initial parameters for a run of the MCMC line-fitting routine. The resulting model is shown along with the spectrum extracted from the nuclear aperture in Figure\;\ref{fig: analysis_and_results: seeing: nuclear_aperture_spectrum}, and its parameters are given in Table\;\ref{tab: muse_f13451_1232: analysis_and_results: seeing: nuclear_model}. The two broadest kinematic components of this model --- henceforth referred to as the `nuclear model' --- are consistent with the `broad' ($1000<\mathrm{FWHM}<2000$\;km\;s$^{-1}$) and `very broad' ($\mathrm{FWHM}\sim3000$\;km\;s$^{-1}$) components of the nuclear [OIII] fits for F13451+1232 presented by \citet{Rose2018}, and correspond to nuclear outflows. The narrowest component of the nuclear spectrum fit is consistent with the `narrow' component of the \citet{Rose2018} [OIII] model, and its kinematics are similar to those of the kiloparsec-scale disk previously detected in CO emission in the object's primary nucleus \citep{Lamperti2022, Holden2024}. Note that while the fit to the nuclear spectrum does not perfectly describe the details of the emission-line profile, it is adequate for the purposes of this study.

\begin{table}
    \renewcommand{\arraystretch}{1.2}
    \centering
    \begin{tabular}{ccc}
	    Peak flux density & Velocity shift & FWHM  \\
	    ($\times10^{-16}$\;erg\;s$^{-1}$\;cm$^{-2}$\;\AA$^{-1}$) & (km\;s$^{-1}$) & (km\;s$^{-1}$) \\
    \hline 
    $3.5\pm0.3$ & $-382\pm44$ & $1006\pm43$  \\
    $1.5\pm0.1$ & $-1248\pm36$ & $2847\pm68$  \\
    $1.1\pm0.4$ & $-20\pm48$ & $413\pm204$  \\
    \end{tabular}
    \caption{Parameters for the kinematic components of the nuclear [OIII] line profile derived from fitting the spectrum extracted from a circular aperture of radius $r=0.4$\;arcseconds around the primary nucleus of F13451+1232. The presented velocity shifts are relative to the galaxy rest frame, and the FWHM values have been corrected for instrumental broadening. The `nuclear model' referred to in this work consists of the two broadest components presented here (the top two rows); the narrowest component (bottom row) is kinematically distinct, and corresponds to the kpc-scale disk in the nucleus of F13451+1232 (see \citealt{Holden2024}).}
    \label{tab: muse_f13451_1232: analysis_and_results: seeing: nuclear_model}
\end{table}

\subsubsection{Emission-line fits of the extended emission}
\label{section: analysis_and_results: seeing: extended_emission_fits}

To determine the large-scale warm-ionised gas kinematics, for each spaxel in the deep datacube, the Bayesian emission-line fitting routine was used to fit a model consisting of the nuclear model (accounting for beam-smeared, nuclear-outflow emission) and $N_\mathrm{g}$ additional Gaussian components (accounting for genuine, non-beam-smeared emission) to the [OIII]$\lambda\lambda$4959,5007 doublet. The centroid wavelengths, widths and relative intensities of the Gaussian components of the nuclear model were fixed, with only the peak flux density being allowed to vary, while all the parameters of the additional Gaussian components were free to vary (allowing them to account for line-profile features not described by the nuclear model). In this way, the line-fitting routine was able to detect and isolate the contributions of genuine extended emission to the line profiles in each spaxel, therefore enabling us to determine the extent to which the nuclear-outflow emission was smeared across the MUSE field of view. Similar approaches to PSF-subtraction and beam-smearing correction have been taken by previous studies (e.g. \citealt{Carniani2015, Kakkad2020, Speranza2024}), although here, only broad components (FWHM\;\textgreater\;500\;km\;s$^{-1}$) were included in the nuclear model.

\subsubsection{Beam smearing of compact outflow emission}
\label{section: analysis_and_results: seeing: psf}

\begin{figure*}
    \begin{subfigure}[t]{0.3355\linewidth}
        \centering
        \includegraphics[width=\textwidth, trim={0 0 1.5cm 0}, clip]{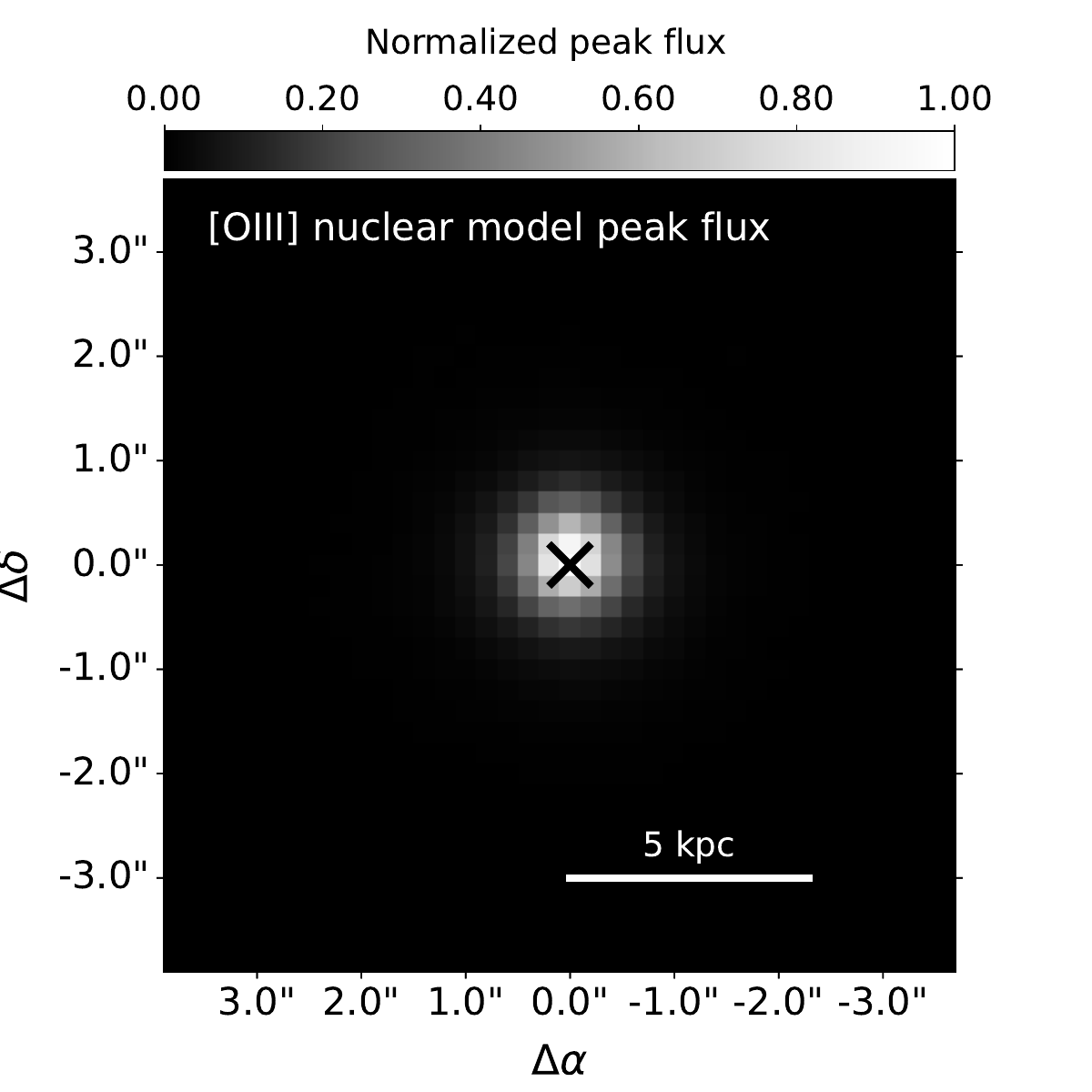}
    \end{subfigure}
    \hspace*{\fill}
    \begin{subfigure}[t]{0.3\linewidth}
        \centering
        \includegraphics[width=\textwidth, trim={2cm 0 1.5cm 0}, clip]{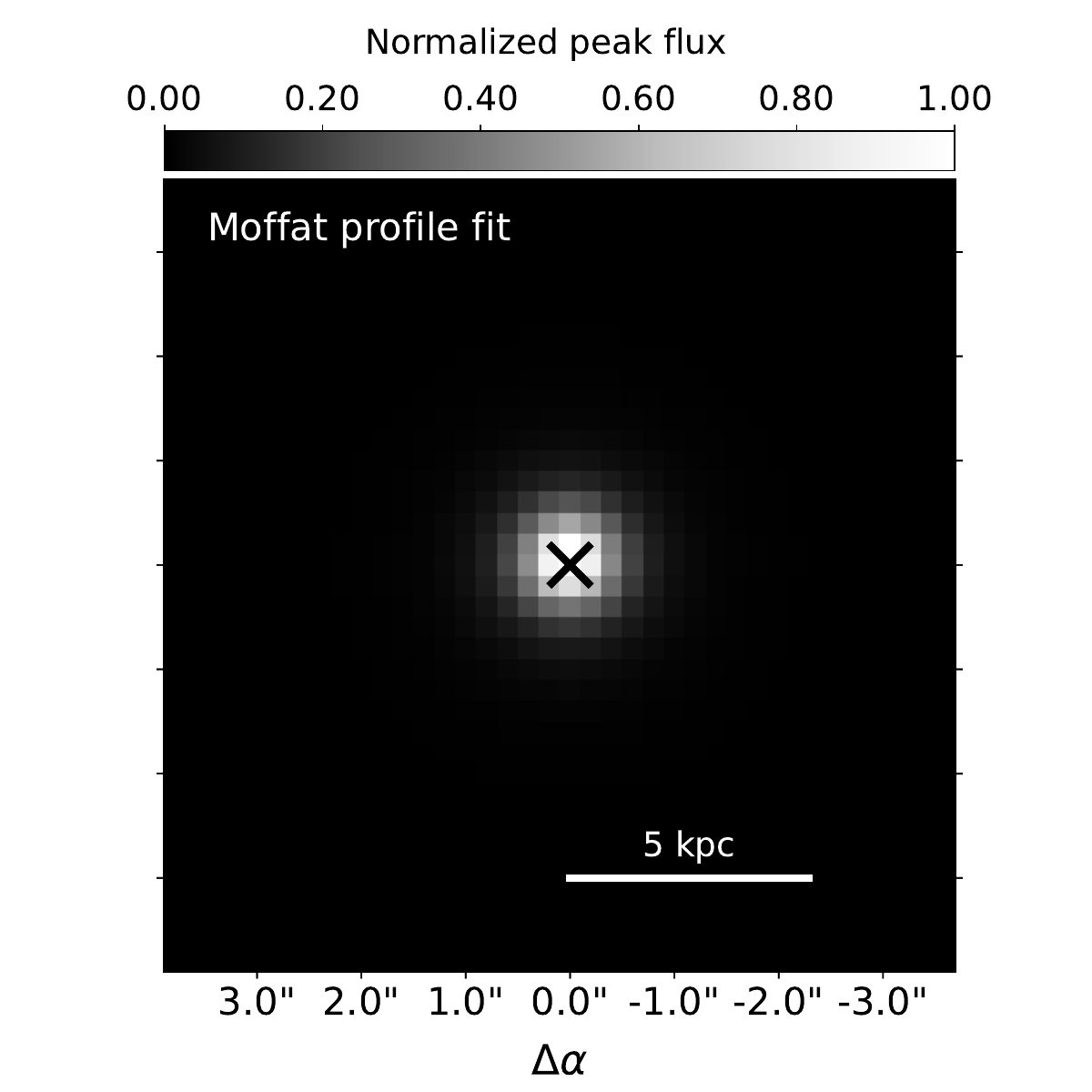}
    \end{subfigure}
    \hspace*{\fill}
    \begin{subfigure}[t]{0.3\linewidth}
        \centering
        \includegraphics[width=\textwidth, trim={2cm 0 1.5cm 0}, clip]{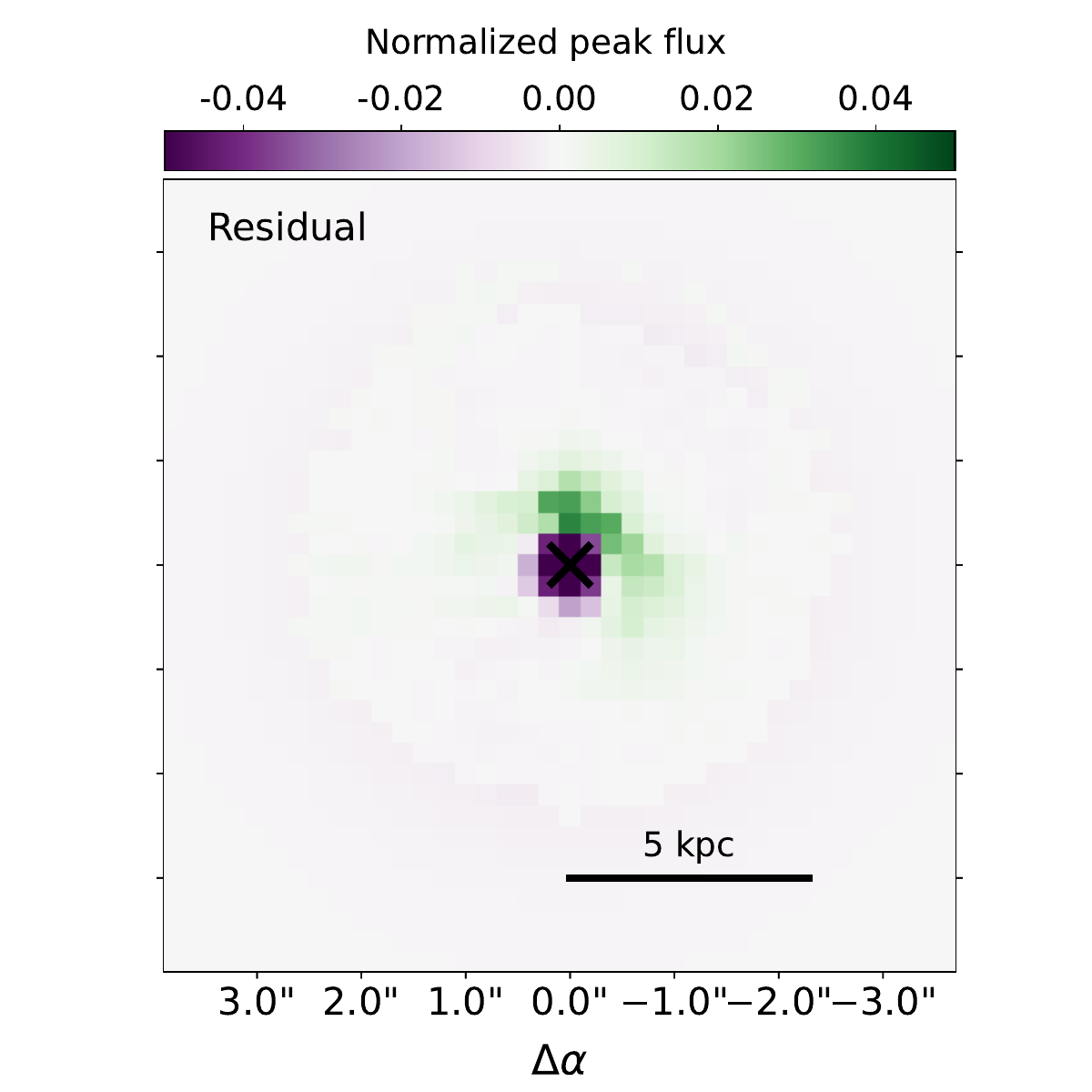}
    \end{subfigure}
    \caption{The spatial distribution of the peak flux of the nuclear model in the [OIII] fits (left panel), the results of two-dimensional Moffat profile fits to this distribution (middle panel), and residuals ([OIII] nuclear model peak flux $-$ model; right panel) in the inner 7\;arcseconds (15\;kpc) of the primary nucleus of F13451+1232 (marked with a black cross). The flux in all cases is normalised to the highest flux value of the nuclear model \textbf{(left panel)}; the full range of normalised flux is used for the left and middle panels, while a significantly reduced range is shown for the residuals (right panel) for presentation purposes. It can be seen that the [OIII]-nuclear-model flux is well-described by a Moffat profile with a corresponding FWHM of $0.74\pm0.02$\;arcseconds, consistent with the seeing value of the dataset (Section \ref{section: observations_and_data_reduction: seeing}).}
    \label{fig: analysis_and_results: seeing: nuclear_model_psf}
\end{figure*}

As a measure of the extent of beam smearing of the compact nuclear-outflow emission, in the left panels of Figure\;\ref{fig: analysis_and_results: seeing: nuclear_model_psf} we present the spatial distribution of the peak flux intensity of the nuclear model in the central $7\times7$\;arcsecond region around the primary nucleus of F13451+1232, as determined by fitting the [OIII]$\lambda\lambda4959,5007$ doublet in each spaxel. The flux of the nuclear-model components appears to be radially symmetric around the location of the primary nucleus, indicative of beam smearing. To investigate this further, a two-dimensional Moffat profile was fit to the spatial flux distribution of the nuclear model, which is shown along with residuals (the flux of the fitted two-dimensional Moffat profile subtracted from the peak flux of the nuclear model) in Figure\;\ref{fig: analysis_and_results: seeing: nuclear_model_psf} --- it can be seen that the spatial distribution of the nuclear-model flux is well described by a Moffat profile with a corresponding FWHM of $0.74\pm0.02$ arcseconds. This value is consistent (within $1\sigma$) with the seeing value measured from the star in the MUSE-DEEP data FOV ($\mathrm{FWHM}_{\star, \mathrm{[OIII]}}=0.79\pm0.10$\;arcseconds: Section\;\ref{section: observations_and_data_reduction: seeing}), providing direct evidence that atmospheric seeing artificially spread emission from the nuclear outflows across the MUSE field of view.

\begin{figure*}
    \hspace*{\fill}
    \begin{subfigure}[]{0.43\linewidth}
        \includegraphics[width=\linewidth, trim={0 1.8cm 0 0}, clip]{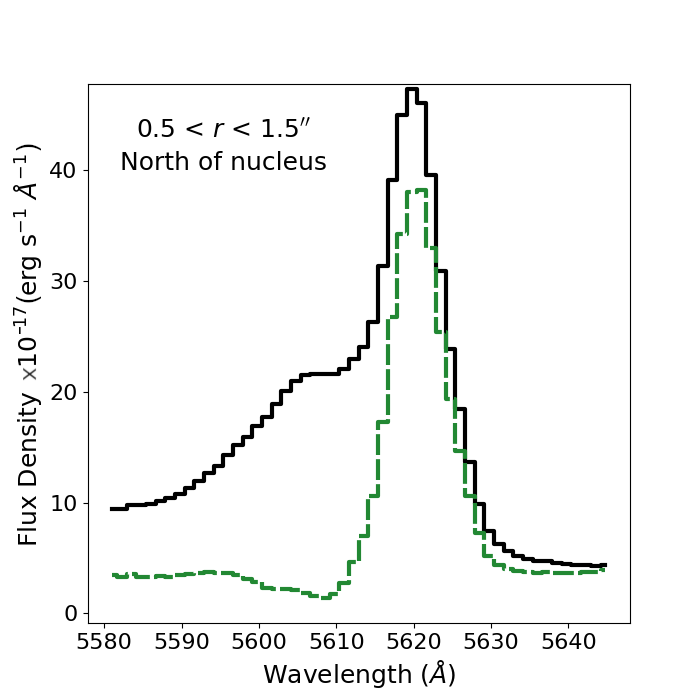}
    \end{subfigure}
    \hspace*{\fill}
    \begin{subfigure}[]{0.41\linewidth}
        \includegraphics[width=\linewidth, trim={1.1cm 1.8cm 0 0}, clip]{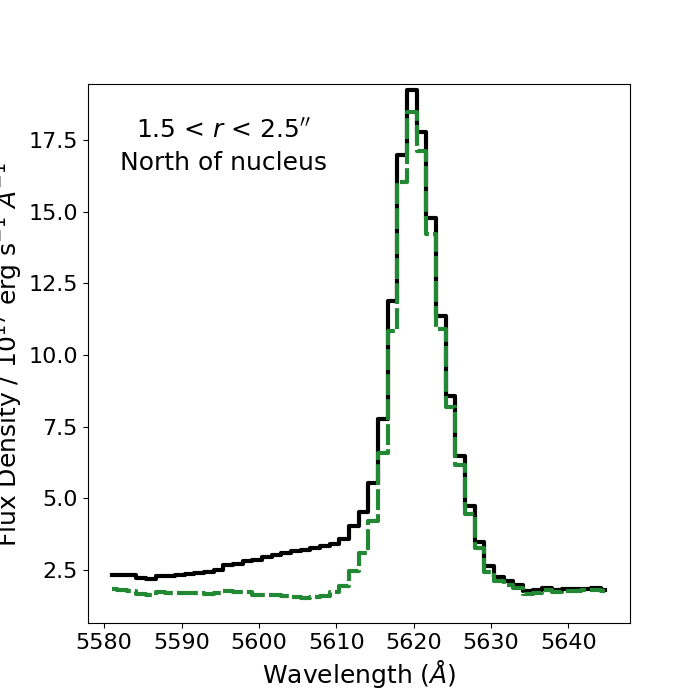}
    \end{subfigure}
    \hspace*{\fill} \\
    \hspace*{\fill}
    \begin{subfigure}[]{0.43\linewidth}
        \includegraphics[width=\linewidth, trim={0 0 0 1.5cm}, clip]{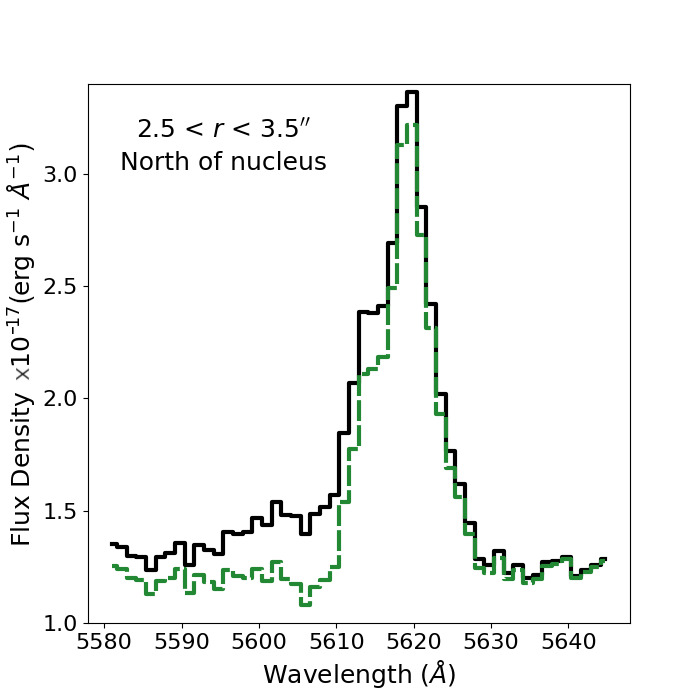}
    \end{subfigure}
    \hspace*{\fill}
    \begin{subfigure}[]{0.41\linewidth}
        \includegraphics[width=\linewidth, trim={1.1cm 0 0 1.5cm}, clip]{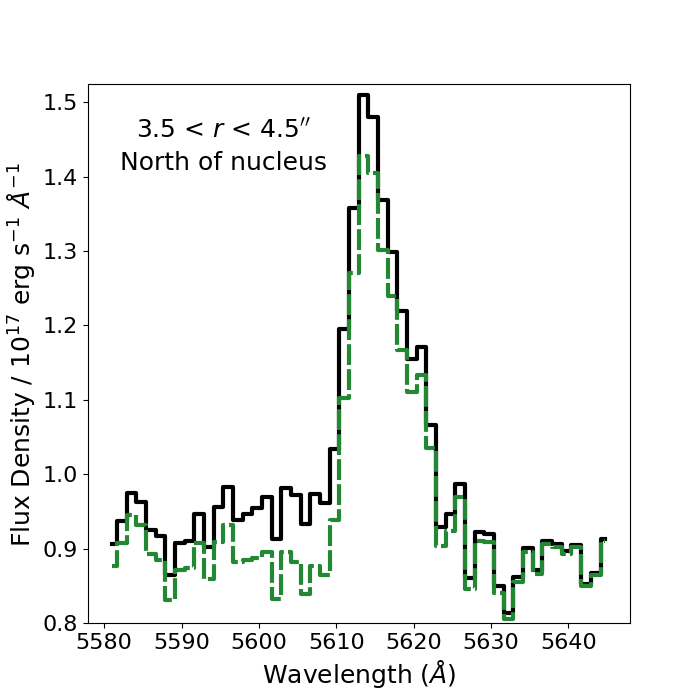}
    \end{subfigure}
    \hspace*{\fill}
    \caption{[OIII]$\lambda5007$ line profile extracted from rectangular $2\times1$\;arcsecond apertures (centred on the primary nucleus in the east-west direction) at increasing radial distances (labelled) to the north of the nucleus, for the original datacube (solid black line) and the datacube with the nuclear model subtracted using a Moffat profile (dashed green line; representing the beam-smearing-corrected case). It can be seen that the beam-smeared nuclear outflow emission is significant between 5595\;\textless\;$\lambda$\;\textless\;5610\;{\AA} ($-1200$\;\textless\;$v$\;\textless\;$-420$\;km\;s$^{-1}$) in all apertures.}
    \label{fig: analysis_and_results: seeing: broadsub_line_profile_comparison}
\end{figure*}

To investigate the residual emission after the beam-smeared nuclear-outflow emission had been accounted for, the [OIII] nuclear model was subtracted from each spaxel of the datacube. This was done by first normalising the peak flux of the nuclear model to unity, multiplying it by the value of the Moffat profile fit to the [OIII] nuclear model flux distribution in each spaxel (middle panels of Figure\;\ref{fig: analysis_and_results: seeing: nuclear_model_psf}; consistent with what is expected from atmospheric seeing, as noted earlier), and subtracting this from the original datacube. To demonstrate the radial extent to which the beam-smeared emission contributes significantly to the line profiles, we extracted a series of rectangular $2\times1$\;arcsecond apertures --- centred on the nucleus in the east-west direction --- at increasing radial distances north of the nucleus from both this datacube and the original datacube. The [OIII]$\lambda5007$ line profiles in these apertures for both datacubes are presented in Figure\;\ref{fig: analysis_and_results: seeing: broadsub_line_profile_comparison}, which demonstrate that the beam-smeared nuclear-outflow emission contributes significantly to the flux between 5595\;\textless\;$\lambda$\;\textless\;5610\;{\AA} ($-1200$\;\textless\;$v$\;\textless\;$-420$\;km\;s$^{-1}$) in all apertures, including in an aperture that covers a radial extent of 3.5\;\textless\;$r$\;\textless\;4.5\;arcseconds ($7.7<r<9.9$\;kpc) north of the primary nucleus. For the spectrum extracted from the aperture closest to the nucleus, there is a slight oversubtraction of flux bluewards of the [OIII]$\lambda$5007 line --- this is a consequence of the small uncertainties in the values for the peak flux and centroid position of the Moffat profile that was subtracted from the datacube.

\subsubsection{Accounting for atmospheric seeing in velocity maps}
\label{section: analysis_and_results: seeing: velocity_maps}

In order to determine the impact of the beam-smeared nuclear-outflow emission on measurements of outflow radial extents and kinematics, the fitting procedure for the original datacube was repeated, but the nuclear model was not included in the fits. This is henceforth referred to as the `free-fitting' case, and was done to provide a test of what would be found if the beam smearing of the nuclear-outflow emission had not been accounted for.

\begin{figure*}
    \hspace*{\fill}
    \centering
    \begin{subfigure}[b]{0.42\linewidth}        
        \includegraphics[width=\textwidth]{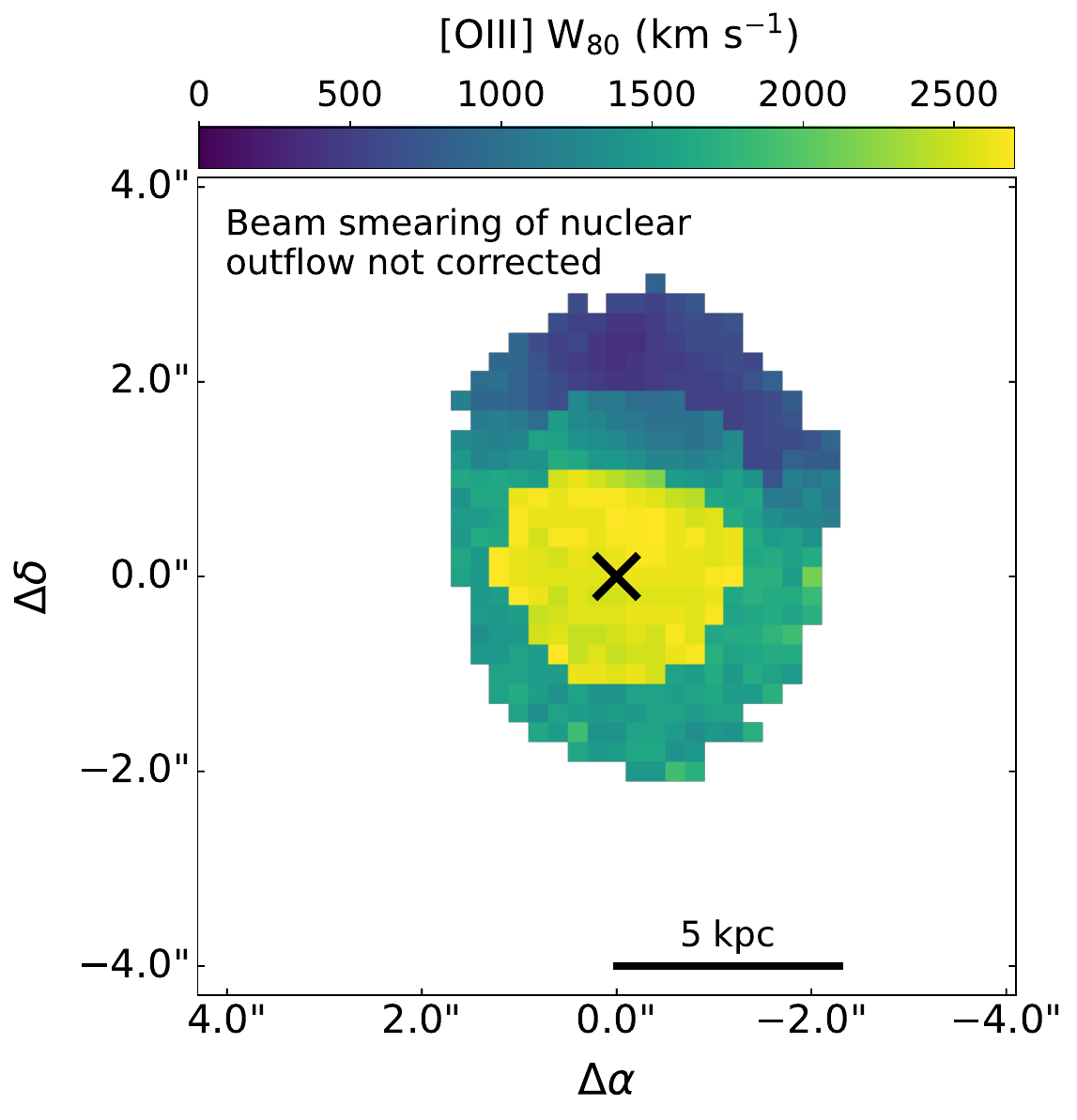}
    \label{fig: analysis_and_results: extended_emission: w80_map_free}
    \end{subfigure}
    \hspace*{\fill}
    \begin{subfigure}{0.373\linewidth}        
        \includegraphics[width=\linewidth, trim={0 0 0 0}, clip]{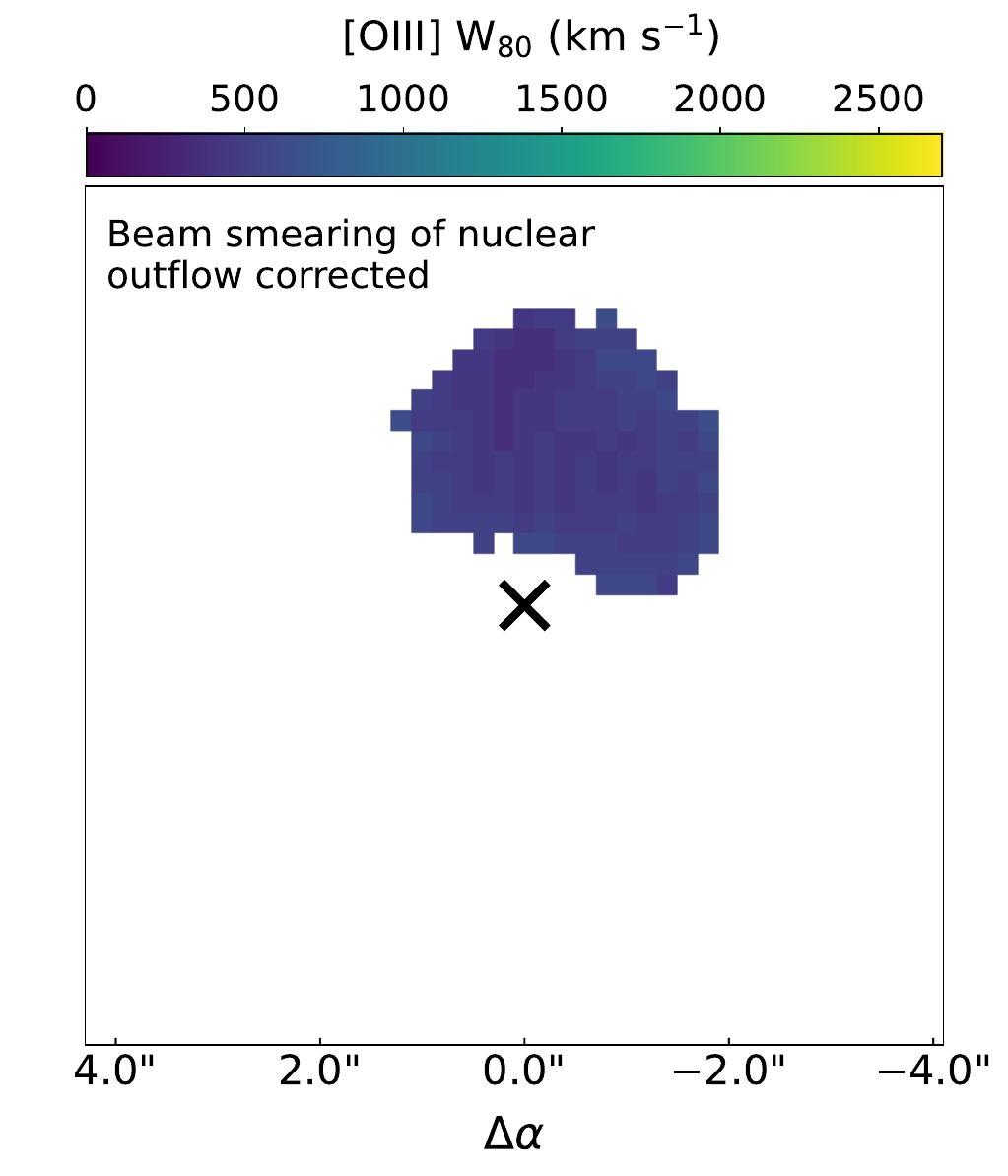}
    \label{fig: analysis_and_results: extended_emission: w80_map_nm}
    \end{subfigure}
    \hspace*{\fill}
    \label{fig: analysis_and_results: extended_emission: w80_maps}
    \vspace*{12pt}
    \centering
    \hspace*{\fill}
    \begin{subfigure}[b]{0.42\linewidth}        
        \includegraphics[width=\textwidth]{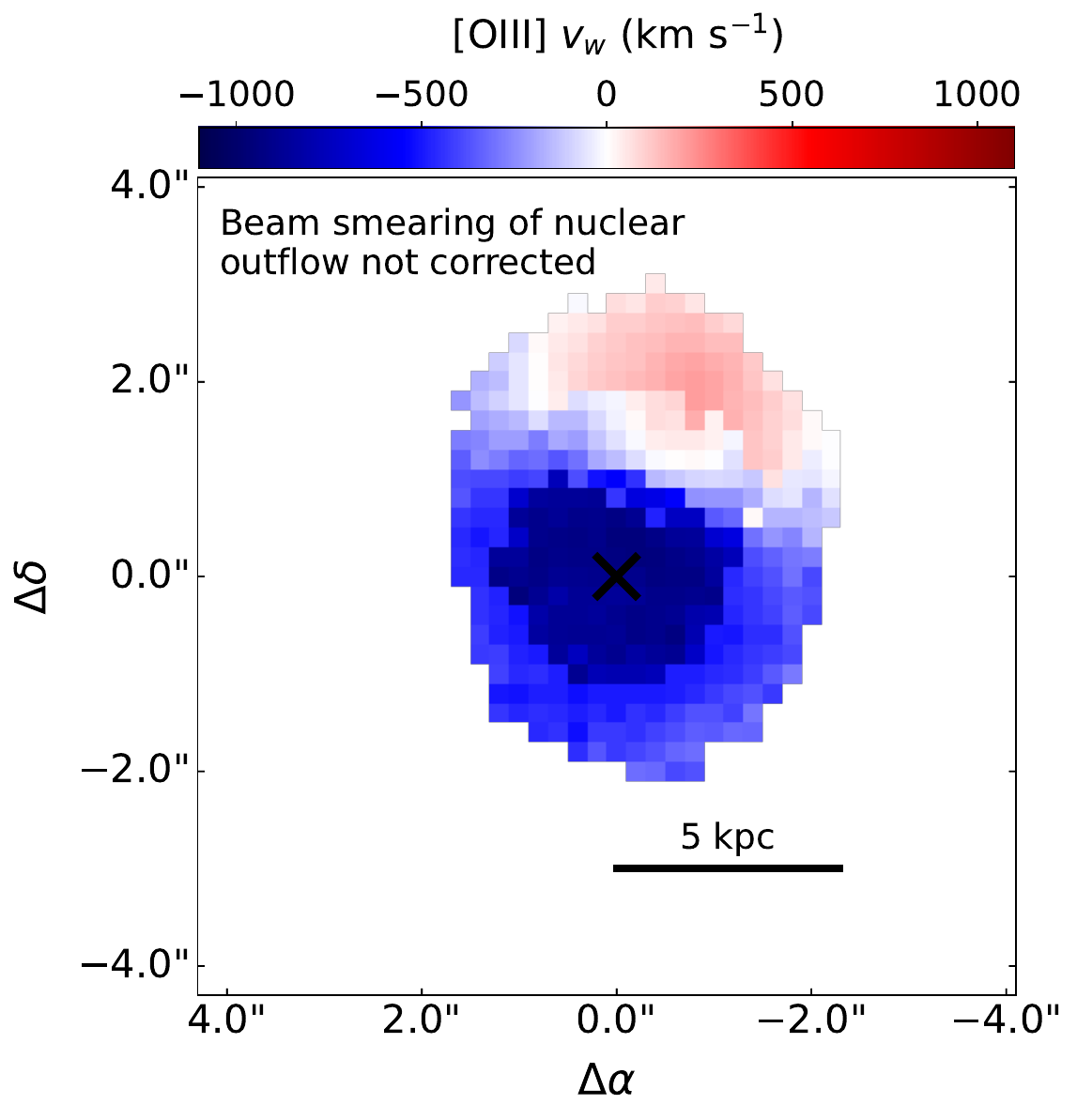}
    \label{fig: analysis_and_results: extended_emission: vw_map_free}
    \end{subfigure}
    \hspace*{\fill}
    \begin{subfigure}{0.373\linewidth}        
        \includegraphics[width=\linewidth, trim={0 0 0 0}, clip]{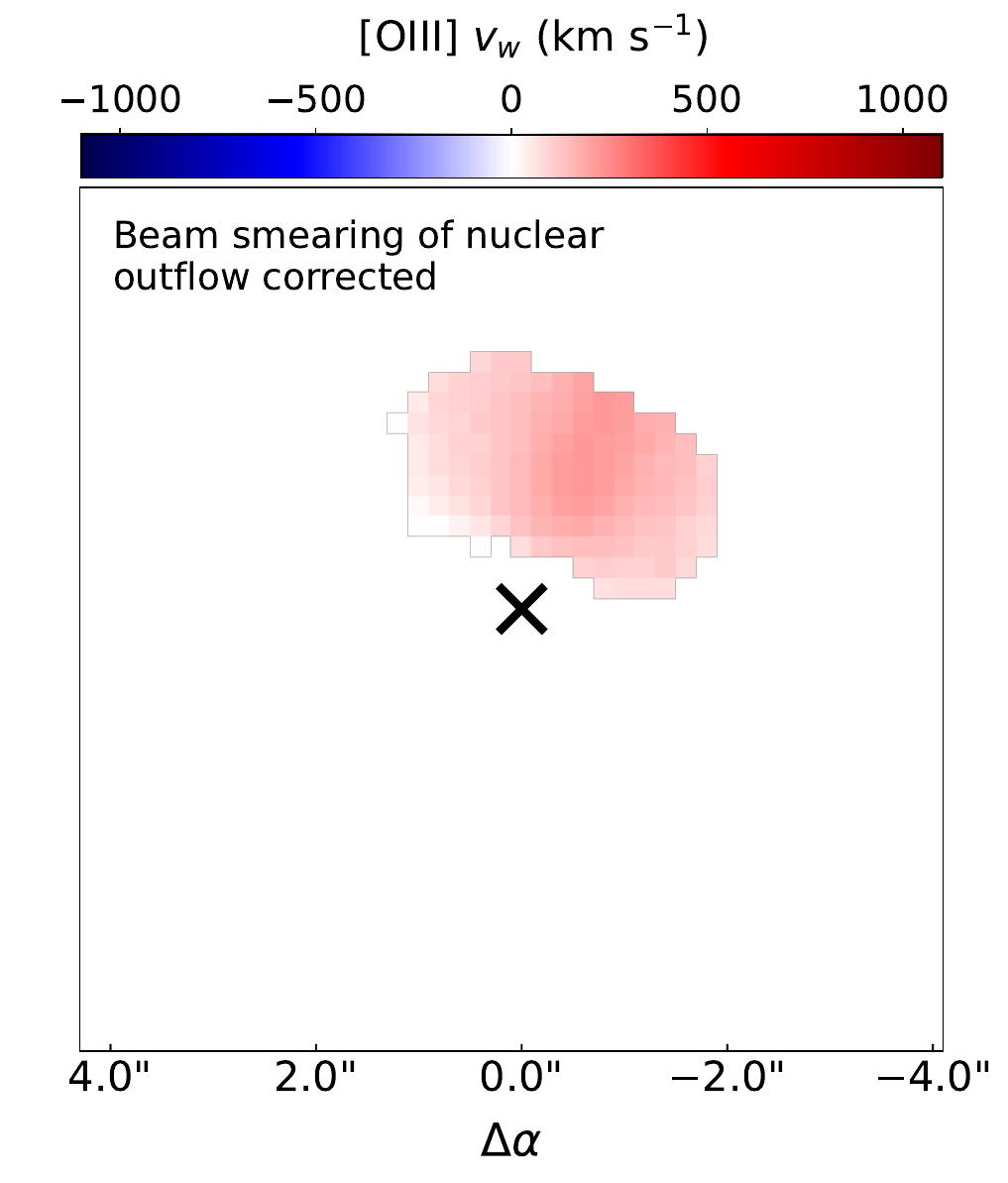}
    \label{fig: analysis_and_results: extended_emission: vw_map_nm}
    \end{subfigure}
    \hspace*{\fill}
    \caption{Non-parametric velocity width ($\mathrm{W_{80}}=v_\mathrm{90}-v_\mathrm{10}$; top panels) and flux-weighted velocity-shift ($v_w$; bottom panels) maps of the central $6\times6$\;arcsecond ($13\times13$\;kpc) region around the primary nucleus of F13451+1232 (black cross), as measured from free-fitting Gaussian components (left panels) and fitting the nuclear model and $N_\mathrm{g}$ Gaussian components (in which only the $N_\mathrm{g}$ Gaussian components were used to measure $W_\mathrm{80}$; right panels). The former case (left panels) is what would be expected had the beam smearing of compact high-velocity outflow emission not been accounted for, while the latter case (right panels) accounts for this beam smearing.}
    \label{fig: analysis_and_results: extended_emission: vw_maps}
\end{figure*}

For the results of both line-fitting approaches, non-parametric [OIII]$\lambda\lambda4959,5007$ velocity-width maps were created. First, any spaxels for which the peak flux density value of the highest-flux Gaussian component was less than $1\sigma_\mathrm{std}$ from the continuum were not considered. In the free-fitting case, all Gaussian components were considered, while in the fits that included the nuclear model, only the additional (non-nuclear-model) Gaussian components were used. For each case, the total fluxes of the Gaussian components involved were calculated, and used to determine the non-parametric 10th- and 90th-percentile velocity shifts ($v_\mathrm{10}$ and $v_\mathrm{90}$) in the galaxy rest frame, which are the velocities that contain 10 and 90\;per\;cent of the emission-line flux, respectively. These percentile velocity shifts were then used to calculate the non-parametric velocity width that contains 80\;per\;cent of the total emission-line flux ($\mathrm{W_{80}}= v_\mathrm{90}-v_\mathrm{10}$) in each spaxel. Furthermore, flux-weighted velocity shifts ($v_w$) in each spaxel for both cases were determined using 
\begin{equation}
	v_w=\frac{\sum_i(F_i\times{v_i})}{\sum_iF_i},
	\label{eq: analysis_and_results: extended_emission: flux_weighted_velocity}
\end{equation}
where $F_i$ and $v_i$ are the fluxes and velocity shifts, respectively, of the individual Gaussian components that constitute the fit to the [OIII]$\lambda\lambda4959,5007$ doublet. The resulting $\mathrm{W_{80}}$ and $v_w$ kinematic maps are presented in Figure\;\ref{fig: analysis_and_results: extended_emission: vw_maps}.

In the velocity maps produced from the free fits (left panels of Figure\;\ref{fig: analysis_and_results: extended_emission: vw_maps}), extreme velocity widths ($\mathrm{W_{80}}\sim2500$\;km\;s$^{-1}$) and shifts ($v_w\sim-1000$\;km\;s$^{-1}$) are seen in a circular region of maximum radial extent $r\sim1$\;arcsecond ($r\sim2.2$\;kpc) centred on the nucleus, in addition to large velocity widths ($\mathrm{W_{80}}\sim1500$\;km\;s$^{-1}$) and shifts ($v_w\sim-500$\;km\;s$^{-1}$) seen in a larger circular region ($r\sim2$\;arcseconds; $r\sim4.4$\;kpc), also centred on the nucleus. The maximum radial extent of detected emission is $r\sim2.5$\;arcseconds ($r\sim5.5$\;kpc) northwest (NW) of the nucleus, where intermediate velocity widths ($\mathrm{W_{80}}\sim500$\;km\;s$^{-1}$) and low velocity shifts ($v_w$\;\textless\;$210$\;km\;s$^{-1}$) are seen. Although this map could be taken as evidence for kinematically-disturbed gas on large scales (up to a radius of $r=2.5$\;arcseconds $=5.5$\;kpc), it is important to note that the effects of beam smearing (Figures \ref{fig: analysis_and_results: seeing: nuclear_model_psf} and \ref{fig: analysis_and_results: seeing: broadsub_line_profile_comparison}) have not been accounted for.

\begin{figure*}
    \hspace*{\fill}
    \centering
    \begin{subfigure}[b]{0.42\linewidth}        
        \includegraphics[width=\textwidth]{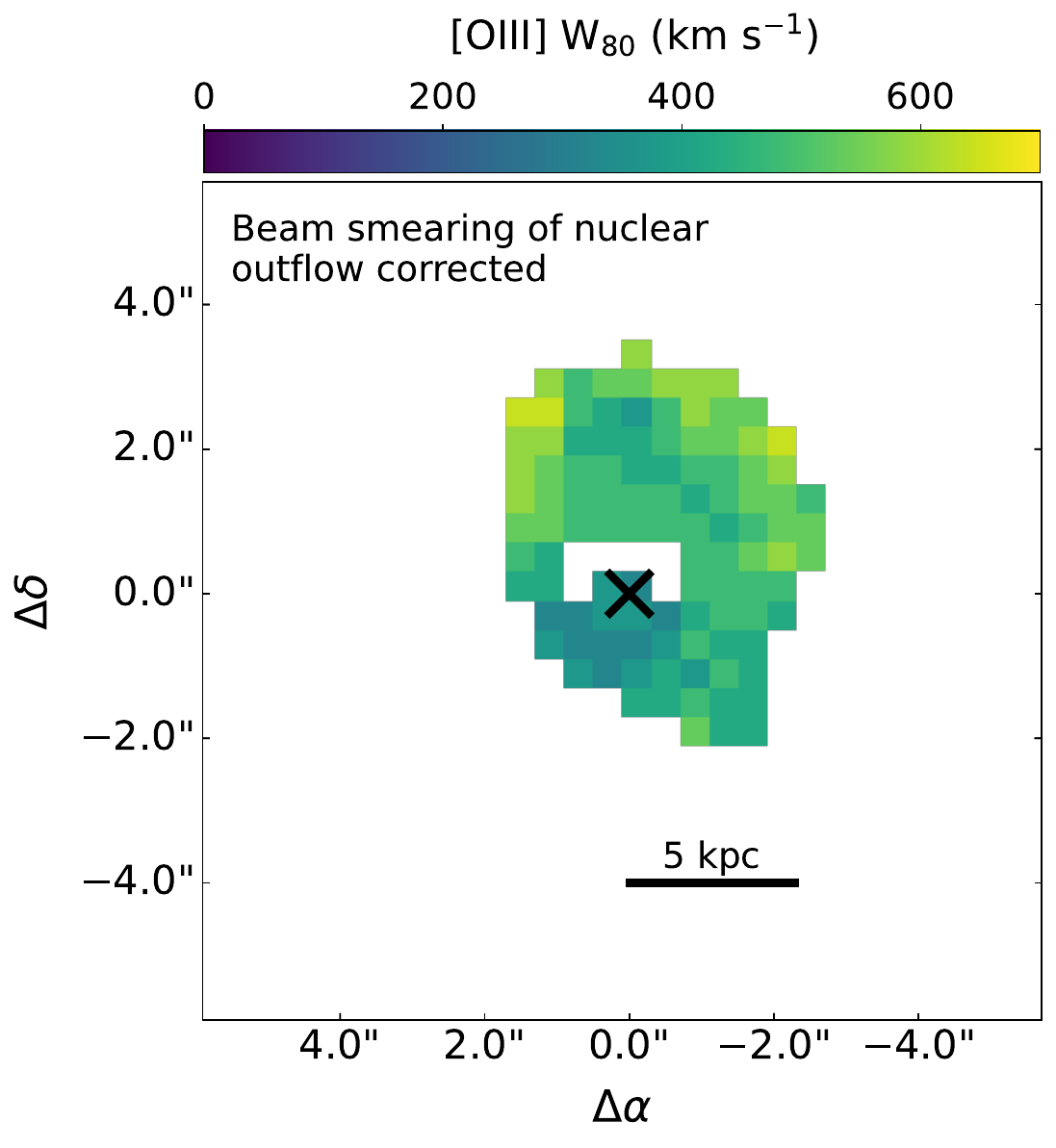}
    \end{subfigure}
    \hspace*{\fill}
    \begin{subfigure}{0.372\linewidth}        
        \includegraphics[width=\linewidth, trim={0 0 0 0}, clip]{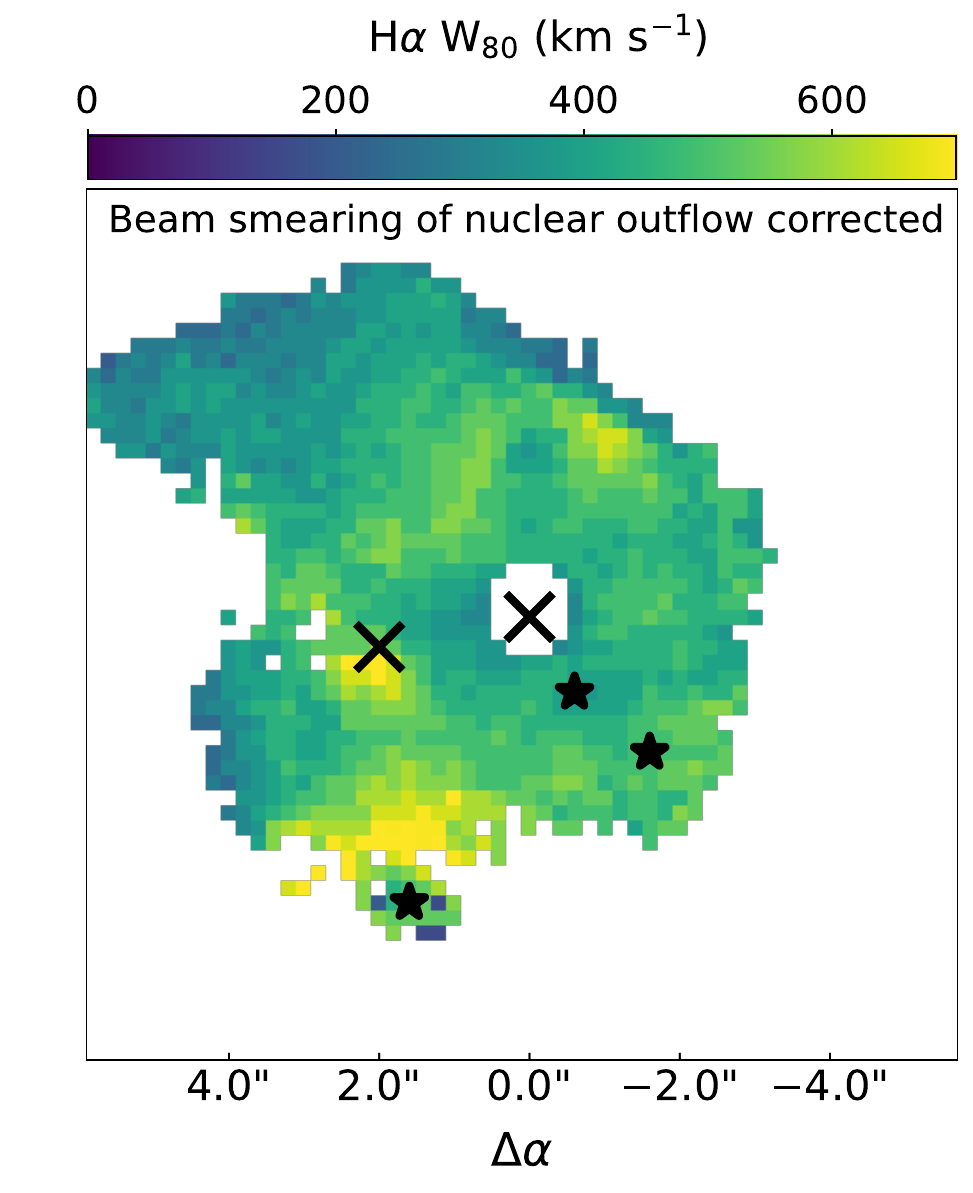}
    \end{subfigure}
    \hspace*{\fill} \\
    \vspace*{12pt}
    \hspace*{\fill}
    \centering
    \begin{subfigure}[b]{0.42\linewidth}        
        \includegraphics[width=\textwidth]{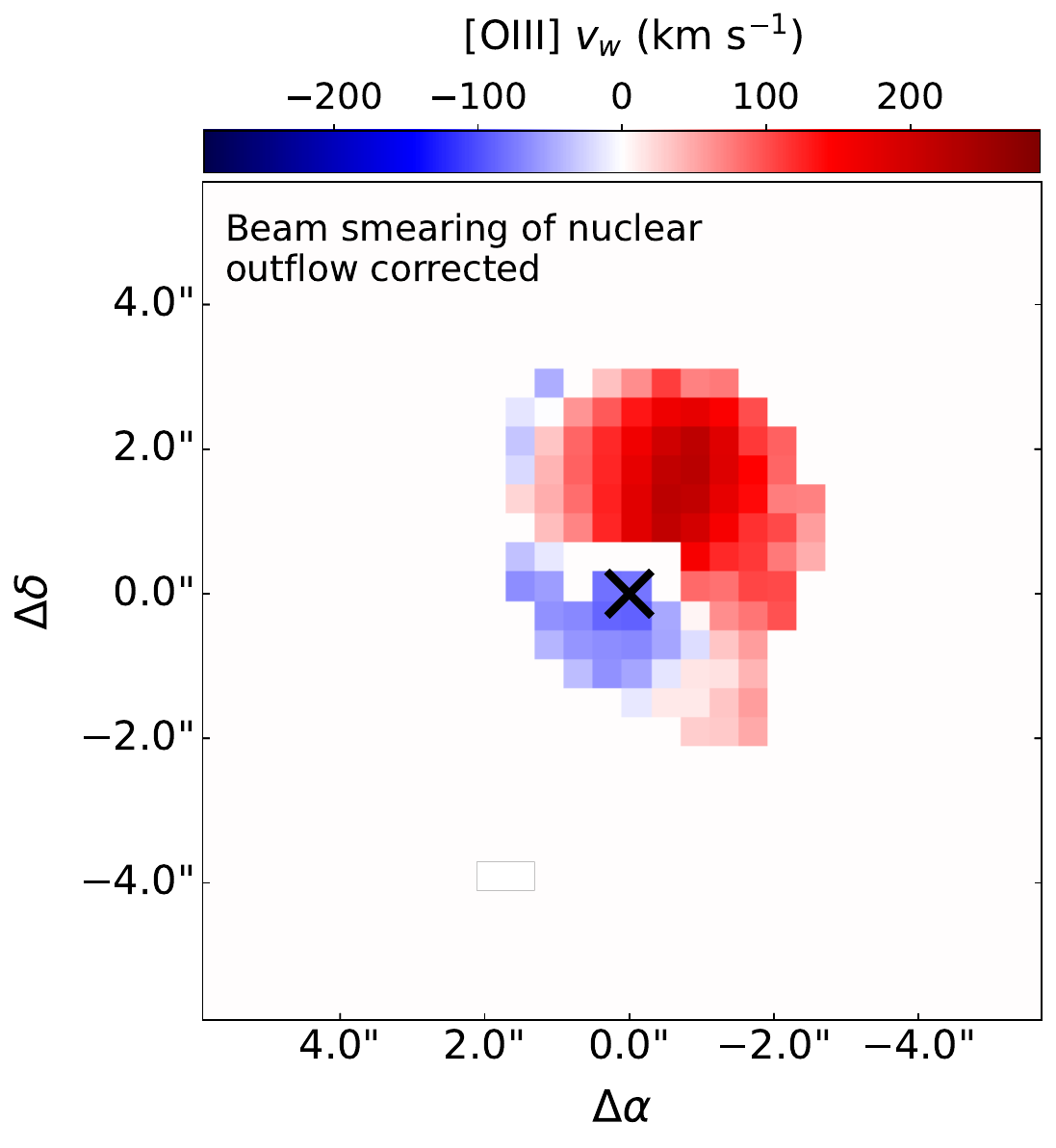}
    \end{subfigure}
    \hspace*{\fill}
    \begin{subfigure}{0.372\linewidth}        
        \includegraphics[width=\linewidth, trim={0 0 0 0}, clip]{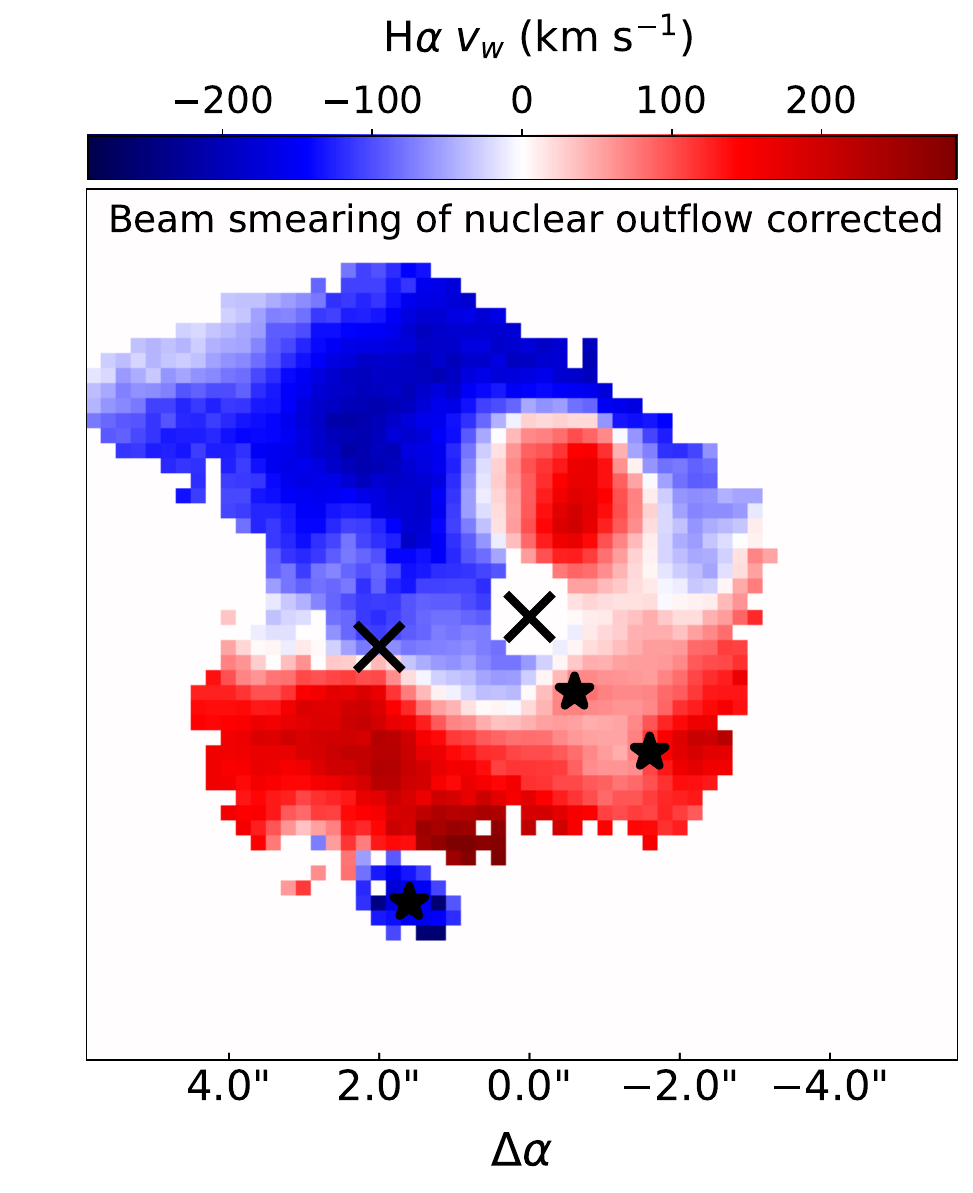}
    \end{subfigure}
    \hspace*{\fill}
    \caption{Non-parametric velocity width ($\mathrm{W_{80}}$: top panels) and flux-weighted velocity shift ($v_w$: bottom panels) maps, centred on the primary nucleus (black cross), for the residual [OIII]$\lambda5007$ (left panels) and H$\alpha$ (right panels) emission in the MUSE-DEEP cube after the PSF of the nuclear-outflow emission had been subtracted. In order to detect lower-surface brightness structures, the residual [OIII] cube was binned by a factor of two before line fitting. These maps reveal that the [OIII] emission resides within a more-spatially-extended region of H$\alpha$ emission, which displays (at least) two distinct kinematic components; the kinematics are mostly modest ($\mathrm{W_{80}}<500$\;km\;s$^{-1}$; $v_w<250$\;km\;s$^{-1}$), with regions of enhanced velocity width (up to $\mathrm{W_{80}}\sim650$\;km\;s$^{-1}$). The locations of the secondary (eastern) nucleus and young star clusters \citep{RodriguezZaurin2007} are marked with a black cross and black star symbols, respectively, in the H$\alpha$ maps (right panels). Note that the spatial and velocity scales are different to those used in Figure\;\ref{fig: analysis_and_results: extended_emission: vw_maps}.}
    \label{fig: analysis_and_results: extended_emission: nuclear_subtracted_vw_maps}
\end{figure*}

By including the nuclear model (Figure\;\ref{fig: analysis_and_results: seeing: nuclear_aperture_spectrum}) in the spaxel fits and only measuring the kinematics of the additional Gaussian components (i.e. those representing genuine, non-nuclear emission), the velocity maps presented in the right panels of Figure\;\ref{fig: analysis_and_results: extended_emission: vw_maps} were produced. Here, the spatially-extended circular region ($r\sim2.5$\;arcseconds) of high velocity widths (1000\;\textless\;$\mathrm{W_{80}}$\;\textless\;2700\;km\;s$^{-1}$) and shifts (350\;\textless\;$\mathrm{W_{80}}$\;\textless\;1000\;km\;s$^{-1}$) produced in the free-fitting case is not seen, as it is accounted for by the nuclear model components. The only location where additional emission is detected is in a region that extends $\sim$2.5\;arcseconds to the NW of the nucleus\footnote{The narrow emission detected in the nuclear aperture (Figure \ref{fig: analysis_and_results: seeing: nuclear_aperture_spectrum}) is not seen in the velocity maps presented in Figure\;\ref{fig: analysis_and_results: extended_emission: vw_maps} as it is too faint in individual spaxels to be detected by the emission-line-fitting routine; it is detected in the velocity maps produced from the cube that has been binned by a factor of two (Figure\;\ref{fig: analysis_and_results: extended_emission: nuclear_subtracted_vw_maps}).}, but the velocity widths are lower ($\mathrm{W_{80}}$\;\textless\;500\;km\;s$^{-1}$) than those seen in the same location in the free-fitting case. Given that, in this region, the emission-line fitting routine required further Gaussian components in addition to the nuclear model to describe the line profiles, we consider this to represent genuine, spatially-extended emission. Conversely, given that the extended region of high-velocity emission seen in the free-fits velocity maps can be accounted for by the nuclear model, and that the peak flux distribution of this model follows the PSF of the seeing disk (Figure \ref{fig: analysis_and_results: seeing: nuclear_model_psf}), we argue that this represents emission from the compact outflows in the nucleus of F13451+1232 that has been artificially spread across the FOV by atmospheric seeing. In this context, it is important to note that this region of high-velocity emission seen in the free-fitting case ($r\sim3$\;arcseconds; left panels of Figure\;\ref{fig: analysis_and_results: extended_emission: vw_maps}) extends far beyond the HWHM of the seeing disk (HWHM$_\star=0.40\pm0.10$\;arcseconds: Section\;\ref{section: observations_and_data_reduction: seeing}), and that, when accounted for, the resulting gas kinematics of any real emission (right panels of Figure\;\ref{fig: analysis_and_results: extended_emission: vw_maps}) are much more modest.

To better characterise the kinematics of the genuinely-extended warm-ionised emission, we spatially binned the nuclear-model-subtracted cube (from which the [OIII] profiles presented in Figure\;\ref{fig: analysis_and_results: seeing: broadsub_line_profile_comparison} were extracted) by a factor of two in both dimensions, and applied the free-fitting approach to the [OIII]$\lambda$5007 line. In this way, we were only considering residual emission after the PSF of the beam-smeared compact outflow emission had been subtracted; the binning allowed for fainter emission to be detected. The resulting flux-weighted velocity shift and $\mathrm{W_{80}}$ maps are presented in the left panels of Figure\;\ref{fig: analysis_and_results: extended_emission: nuclear_subtracted_vw_maps}. Using this approach, the genuinely-extended [OIII] emission to the NE of the nucleus (previously seen in Figure\;\ref{fig: analysis_and_results: extended_emission: vw_maps}) is detected, in addition to previously-unseen emission at the position of the nucleus and to the south, east, and northeast. The bulk of this emission has line widths that are consistent with those seen in the previous beam-smearing-corrected velocity maps ($\mathrm{W_{80}}<500$\;km\;s$^{-1}$), however there are regions of enhanced velocity width (up to $\mathrm{W_{80}}\sim500$\;km\;s$^{-1}$). The velocity shifts of the detected emission are consistent with that seen in the previous velocity maps, while also revealing blueshifts of up to $|v_w|<150$\;km\;s$^{-1}$ at the position of the nucleus.

Since the H$\alpha$ emission line is considerably brighter at large radial distances from the AGN in the MUSE-DEEP cube than the [OIII]$\lambda\lambda4959,5007$ doublet, we repeated the PSF-subtraction method for the H$\alpha$+[NII] blend to quantify any larger-scale kinematics of the warm-ionised gas. First, we fit the H$\alpha$+[NII] blend in each spaxel using a nuclear model and $N_g$ Gaussian profiles, as was done for [OIII]$\lambda\lambda4959,5007$ in Sections \ref{section: analysis_and_results: seeing: nuclear_aperture_extraction} and \ref{section: analysis_and_results: seeing: extended_emission_fits} --- the nuclear H$\alpha$ and {[}NII{]}$\lambda\lambda$6548,6584 line profiles were significantly blended, and so the nuclear model consisted of a single broad ($\mathrm{FWHM}>1000$\;km\;s$^{-1}$) Gaussian for each of the lines (with equal line widths, and wavelength separations and [NII]-flux ratio taken from atomic physics) and $N_\mathrm{g}$ Gaussian components for the blend of the three lines. Next, we modelled the PSF of the nuclear H$\alpha$+[NII] emission by fitting a Moffat function to the flux distribution of the H$\alpha$+[NII] nuclear model across the field of view, and subtracted this from the original, unbinned datacube (following the methodology outlined in Section \ref{section: analysis_and_results: seeing: psf}). The corresponding FWHM of the Moffat-profile fit was $\mathrm{FWHM_{H\alpha}}=0.43\pm0.01$\footnote{The smaller FWHM value for the seeing disk measured at the wavelength of H$\alpha$ compared to that measured at the wavelength of the [OIII]$\lambda\lambda4959,5007$ doublet is consistent with what is expected from the wavelength-dependence of atmospheric seeing and the MUSE-WFM-AO PSF \citep{Fusco2020}.}; fitting a Moffat profile to the star in the field of view of the observations at the wavelength of H$\alpha$ gave $\mathrm{FWHM_{\star,H\alpha}}=0.44\pm0.02$, thus providing further evidence that nuclear-outflow emission had been smeared by atmospheric seeing. Finally, we applied the free-fitting approach to the residual H$\alpha$+[NII] emission, from which the kinematic maps presented in the bottom panels of Figure\;\ref{fig: analysis_and_results: extended_emission: nuclear_subtracted_vw_maps} were created. These beam-smearing-corrected H$\alpha$ velocity maps reveal emission that extends to radial distances of $r\sim5.5$\;arcseconds ($r\sim12.4$\;kpc), with two clear components in velocity shift: blueshifted emission to the northeast, and redshifted emission to the west and southwest. In the regions where both [OIII]$\lambda\lambda4959,5007$ and H$\alpha$+[NII] are detected, the kinematics derived from both lines are consistent, including the regions of enhanced $\mathrm{W_{80}}$. Overall, by subtracting the nuclear-model from the datacube and fitting the residual emission, we find that the [OIII] emission resides within an extended, lower-ionisation region (traced by H$\alpha$) that has mostly modest velocity widths ($\mathrm{W_{80}}<500$\;km\;s$^{-1}$) with regions of enhanced velocity width (up to $\mathrm{W_{80}}\sim650$\;km\;s$^{-1}$) and two clearly-separated blue- and redshifted components that extend to large scales ($r\sim5.5$\;arcseconds; $r\sim12.4$\;kpc).

\subsection{The nature of the extended gas kinematics in F13451+1232}
\label{section: analysis_and_results: kinematics}

The large-scale kinematics measured here from the MUSE-DEEP data are consistent with those that have already been established for F13451+1232 by previous studies \citep{Holt2003, RodriguezZaurin2007, Perna2021,Perna2022}: complex velocity structures with no clear evidence for ordered rotation on $r\sim5$--20\;kpc scales and with the bulk of the gas showing maximum (projected) flux-weighted-velocity-shift amplitudes of $|v_w|\sim200$--300\;km\;s$^{-1}$. Such properties are typical of nearby ULIRGs \citep[e.g.][]{Colina2005, Westmoquette2012,Perna2022}, and are consistent with the disordered motions that are expected in the final stages of a galaxy merger. However, larger velocity shifts for F13451+1232 --- up to $v\sim500$\;km\;s$^{-1}$ relative to the galaxy rest frame and $v\sim$300--550\;km\;s$^{-1}$ relative to the local halo gas --- are also observed in distinct kinematic components at the sites of massive, young star clusters (Figure\;\ref{fig: analysis_and_results: extended_emission: nuclear_subtracted_vw_maps}): a pair of clusters located $\sim$10\;kpc to the southwest of the primary nucleus, and a single cluster $\sim$5.5\;kpc to the southeast of the western (primary) nucleus (see \citealt{RodriguezZaurin2007}). At these locations, the velocity widths of the narrowest components ($\mathrm{FWHM}<150$\;km\;s$^{-1}$) are below those that are measured for much of the low-ionisation halo ($300<\mathrm{FHWM}<500\;$\;km\;s$^{-1}$), and the emission-line-flux ratios of the narrow components are consistent with photoionisation by stars, in contrast to the LINER (low ionisation nuclear emission region)-like ionisation observed elsewhere in the halo (\citealt{RodriguezZaurin2007}; see also Chapter\;5 of \citealt{SpenceThesis}). 

Correcting for the beam-smearing of the nuclear-outflow emission in F13451+1232 allows us to examine the emission-line gas kinematics at smaller radial distances from the nucleus than past studies ($r<5$\;kpc) and search for signs of AGN-driven outflows. Interestingly, the H$\alpha$ kinematics show moderately-broad velocity widths ($300<\mathrm{W_{80}}<500$\;km\;s$^{-1}$) across the entire central region (shown in Figure\;\ref{fig: analysis_and_results: extended_emission: nuclear_subtracted_vw_maps}). Such uniformly-broad lines would not be expected for a settled disk that is undergoing regular gravitational motion\footnote{The maximum possible velocity width measured for a line-of-sight passing through a disk that is seen edge-on cannot be greater than the disk rotation velocity, assuming a flat rotation curve. If such a disk existed in F13451+1232 on kiloparsec scales and had a rotation velocity equivalent to that of the molecular disk observed in CO(1-0) emission on $<$1\,kpc scales \citep{Lamperti2022, Holden2024}, the expected measured value of $\mathrm{W_{80}}$ would be less than 300\,km s$^{-1}$.}. Indeed, while there is an overall trend from blueshift to redshift moving from north to south, the observed velocity structure is not consistent with regular rotation: the boundary between blue- and redshifted regions does not pass through the primary nucleus, and there is a region of redshifted emission located $\sim$1\;arcsec ($\sim$2.2\;kpc) to the north of the primary nucleus that likely represents the emission-line arc structure visible in narrow-band [OIII] HST imaging \citep{Tadhunter2018}.

A closer inspection of the extended gas kinematics reveals regions with relatively-large velocity widths ($500<\mathrm{W_{80}}<650$\;km\;s$^{-1}$) that might, by some criteria (e.g \citealt{Spence2016, Venturi2023}), be considered to indicate the presence of AGN-driven outflows. While in some of these regions the broad lines are likely due to the superposition of red- and blue-shifted kinematic components and the effects of beam smearing, this is not always the case. In particular, a region located $\sim$1\;arcsecond north of a pair of star clusters to the southwest (shown as a red dash-dotted rectangle in Figure\;\ref{fig: analysis_and_results: extended_emission: apertures}) presents a highly complex emission-line profile (Figure \ref{fig: analysis_and_results: kinematics: s_broad_region_halpha}): as well as the beam-smeared narrow component associated with star clusters, there is a much broader component ($W_{80}\sim550$\;km\;s$^{-1}$) that can be modelled as the superposition of two Gaussian components (each of $\mathrm{FWHM}\sim350$\;km\;s$^{-1}$) that are separated by $v\sim280$\;km s$^{-1}$. 

While it is difficult to completely rule out that such complex emission-line kinematics represent AGN-driven outflows, perhaps a more natural explanation is that they simply reflect the disordered gravitational motions and dissipational settling of gas that would be expected in the final stages of a galaxy merger. In this context, it is notable that on smaller ($<1$\;kpc) radial scales from the AGN, where the dynamical timescales are shorter, a regularly-rotating molecular gas disk has been detected around the western nucleus using high-resolution CO(1-0) observations, in addition to a fast molecular outflow \citep{Holden2024}. Therefore, overall, we favour the scenario that the large-scale warm-ionised-gas kinematics in F13451+1232 are due to gravitational motions in a galaxy merger, rather than representing AGN-driven outflows.

\subsection{Energetics of the extended warm-ionised emission}
\label{section: analysis_and_results: extended_emission}

Although we interpret the large-scale gas kinematics in F13451+1232 as being due to gravitational motions in the ongoing galaxy merger, in order to determine the impact that this gas would have on the host galaxy if it was actually outflowing, in this section we derive upper limits for its masses, mass outflow rates, and kinetic powers; we also investigate the extent to which beam smearing of the nuclear-outflow emission may affect these properties.

\subsubsection{Aperture extraction from the MUSE-DEEP datacube}
\label{section: analysis_and_results: extended_emission: apertures}

To ensure that the signal of important diagnostic emission lines (and the potential contamination from the beam-smeared nuclear emission) was sufficient for robust characterisation, rectangular apertures of various sizes and positions in the MUSE-DEEP field-of-view were selected. The spaxels contained in a given aperture were summed to give a single one-dimensional spectrum, and the flux errors of each spaxel were added in quadrature.

\begin{figure*}
    \centering
    \includegraphics[width=\linewidth, trim={0cm 0 0 0cm}, clip]{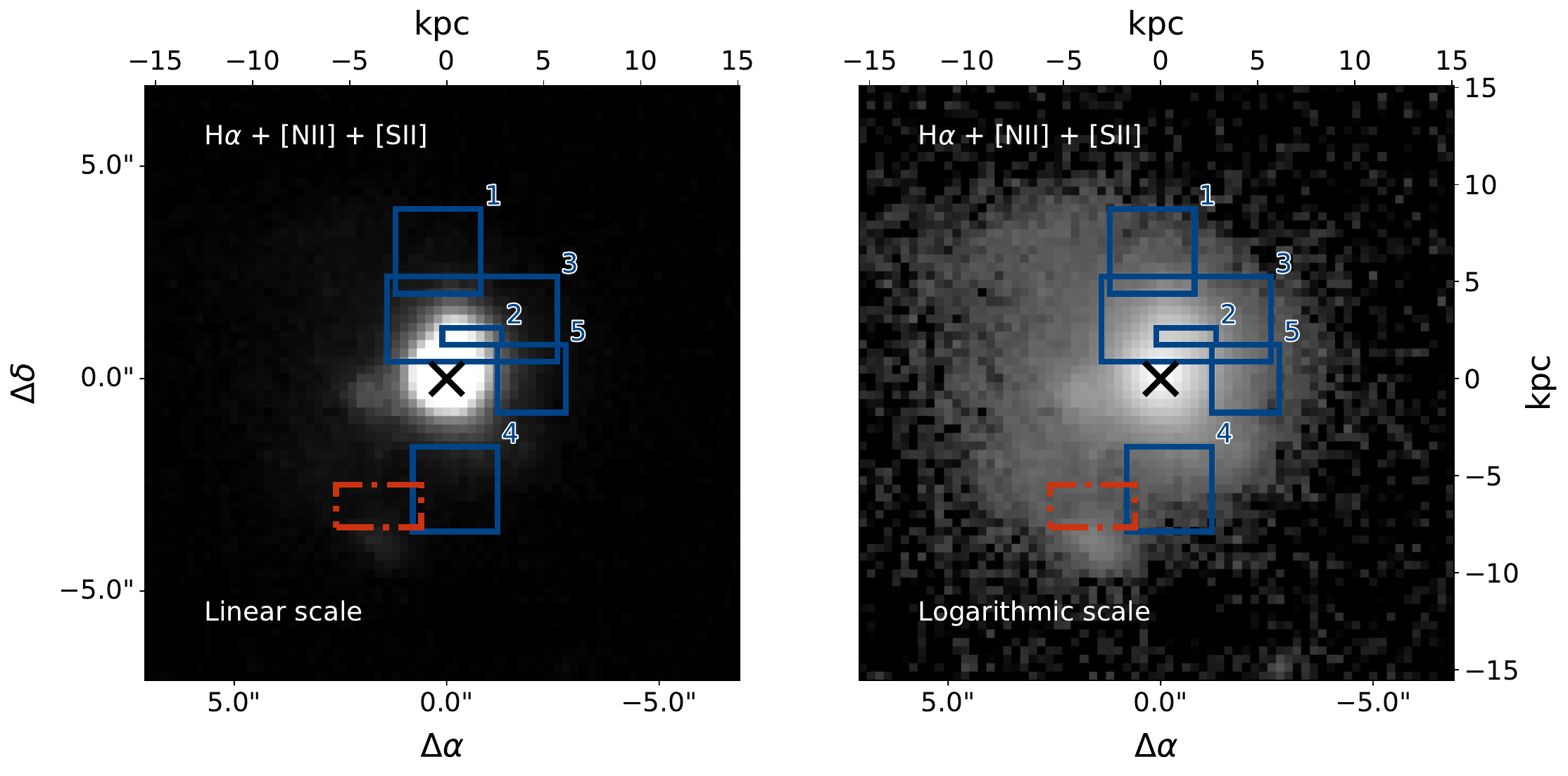}
    \caption{Continuum-subtracted H$\alpha$ + [NII]$\lambda\lambda$6548,6584 + [SII]$\lambda\lambda$6717,6731 image (left: linear scale; right: logarithmic scale) of F13451+1232 with the locations and sizes of the selected apertures shown with solid blue lines; the dash-dotted red rectangle shows the region north of a pair of star clusters that was extracted to produce the line profile presented in Figure\;\ref{fig: analysis_and_results: kinematics: s_broad_region_halpha}. The aperture numbers are labelled at the top right of each, and the position of the primary nucleus is marked with a black cross.}
    \label{fig: analysis_and_results: extended_emission: apertures}
\end{figure*}

\begin{figure}
    \centering
    \vspace*{1cm}
    \includegraphics[width=1\linewidth]{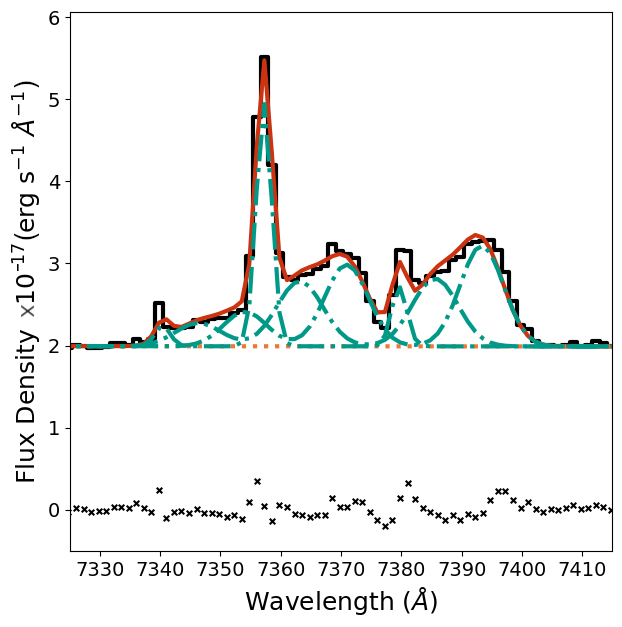}
    \caption{H$\alpha$+[NII] line profile extracted from a region (shown as a dash-dotted red rectangle in Figure\;\ref{fig: analysis_and_results: extended_emission: apertures}) north of a pair of known star clusters (see \citealt{RodriguezZaurin2007}). A beam-smeared narrow component associated with the clusters can be seen, in addition to a redshifted broad component (which is well-described by two Gaussian components).}
    \label{fig: analysis_and_results: kinematics: s_broad_region_halpha}
\end{figure}

The locations and sizes of the apertures --- shown in Figure\;\ref{fig: analysis_and_results: extended_emission: apertures} --- were chosen to cover distinct flux structures seen in H$\alpha$+[NII]$\lambda\lambda$6548,6584+[SII]$\lambda\lambda$6717,6731 images produced from the MUSE-DEEP cube and high-spatial-resolution (0.05\;arcseconds\;pix$^{-1}$) [OIII] imaging presented by \citet{Tadhunter2018}. Moreover, the sizes of the apertures were also chosen to include sufficient signal in the required diagnostic lines. Aperture\;1 was selected as it was the furthest location (in radial distance) to the north of the primary nucleus in which the fainter emission lines were well-detected; Aperture 2 was chosen to cover the arc-like structure that is located $\sim$1\;arcsecond to the NW of the nucleus that is seen in the high-resolution [OIII] HST imaging by \citet{Tadhunter2018}; Aperture 3 covers the approximate area of genuinely-extended [OIII] emission seen in the velocity maps (right panels of Figure\;\ref{fig: analysis_and_results: extended_emission: vw_maps}), and Apertures 4 and 5 were placed at the furthest radial distances to the south and west respectively that contained sufficient signal in the required diagnostic lines. We did not extract an aperture covering the secondary nucleus ($r\sim2$\;arcseconds east of the primary nucleus) because the [OIII]$\lambda\lambda4959,5007$ doublet at this location has a low equivalent width (EW), is dominated by beam-smeared emission from the primary nucleus, and the underlying continuum is complex. Likewise, the region that presents bright H$\alpha$ + {[}NII{]}$\lambda\lambda$6548,6584 + {[}SII{]}$\lambda\lambda$6717,6731 emission 3.6\;arcseconds to the southeast of the primary nucleus (dash-dotted red rectangle in Figure\;\ref{fig: analysis_and_results: extended_emission: apertures}; see the extracted H$\alpha$+[NII] line profile in Figure\;\ref{fig: analysis_and_results: kinematics: s_broad_region_halpha}) --- corresponding to the location of a known HII region in this object \citep{RodriguezZaurin2007} --- did not contain sufficient signal in the [OIII] lines for measurement.

\subsubsection{Aperture emission-line fits}
\label{section: analysis_and_results: extended_emission: aperture_line_fits}

For each aperture, the line profiles of the [OIII]$\lambda\lambda$4959,5007 doublet (used for kinematics), the H$\alpha$ + {[}NII{]}$\lambda\lambda$6548,6584 blend (the former used to determine gas masses), and the {[}SII{]}$\lambda\lambda$6717,6731 doublet (used to determine electron densities) were fit with two methods: one using their respective nuclear models and $N_\mathrm{g}$ Gaussian components, and the other using only $N_\mathrm{g}$ Gaussian components (free-fitting), mirroring the approach taken in Section \ref{section: analysis_and_results: seeing}. Regarding the nuclear-model fits, in the case of the [OIII]$\lambda\lambda$4959,5007 doublet, the nuclear model described in Section\;\ref{section: analysis_and_results: seeing: nuclear_aperture_extraction} (Figure\;\ref{fig: analysis_and_results: seeing: nuclear_aperture_spectrum}; Table\;\ref{tab: muse_f13451_1232: analysis_and_results: seeing: nuclear_model}) was used, and for H$\alpha$+[NII], the nuclear model described in Section \ref{section: analysis_and_results: seeing: velocity_maps} was used. For the {[}SII{]}$\lambda\lambda$6717,6731 doublet, a nuclear model was defined using a similar method to that for [OIII]: $N_\mathrm{g}$ Gaussian components were used for each line; constraints from atomic physics \citep{Osterbrock2006} were applied to fix the wavelength separation of the lines, and the widths of the lines were set to be equal for a given component. However, in this case, the ratio of the peak flux densities of the lines was required to be in the range 0.44\;\textless\;[SII](6717/6731)\;\textless\;1.45.

\begin{figure*}
    \centering
    \begin{subfigure}[t]{0.9\linewidth}
        \begin{minipage}{0.48\linewidth}
            \centering
            \textbf{Nuclear-model fits}
        \end{minipage}
        \hfill
        \begin{minipage}{0.42\linewidth}
            \centering
            \textbf{Free fits}
        \end{minipage}
        \vfill
        \includegraphics[width=0.45\textwidth]{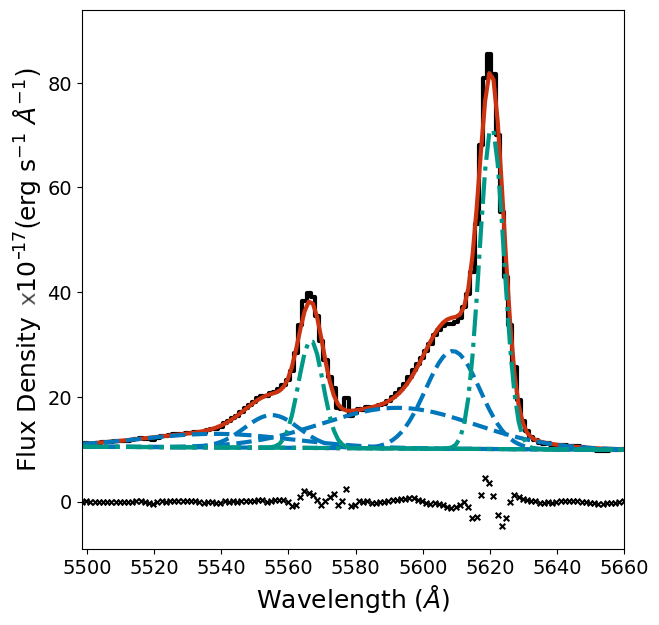}
	\hspace{1.22cm}
        \includegraphics[width=0.45\textwidth]{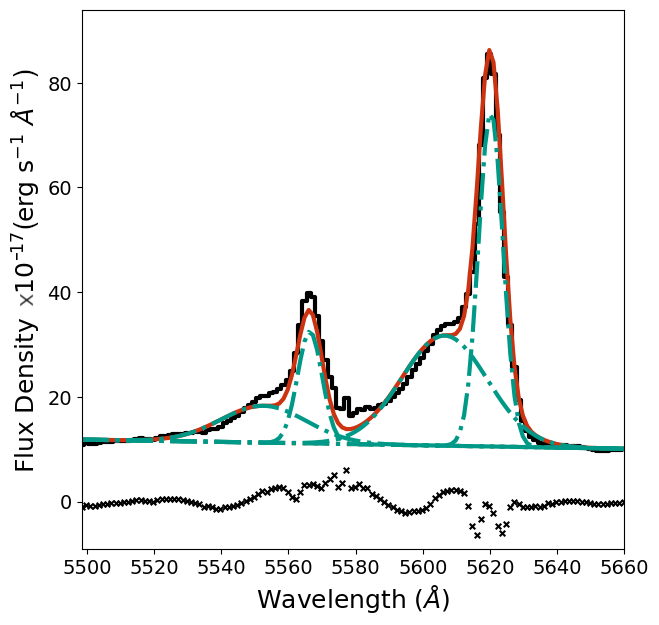}
        \vfill
        \includegraphics[width=0.435\textwidth]{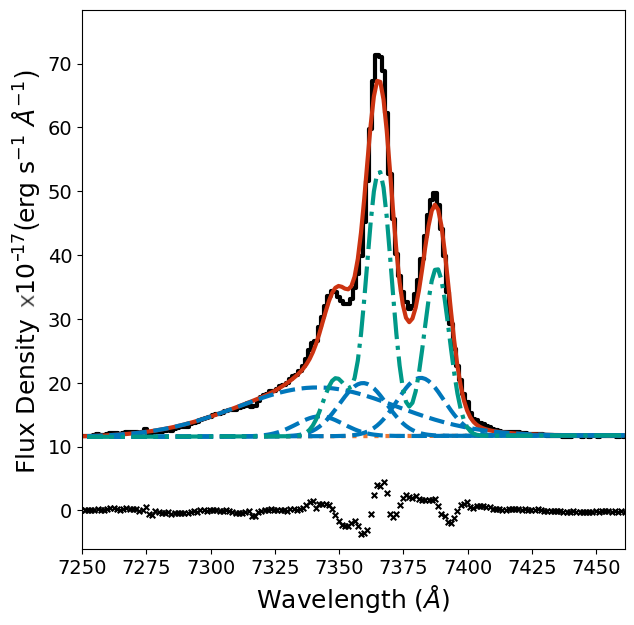}
        \hspace{1.42cm}
        \includegraphics[width=0.435\textwidth]{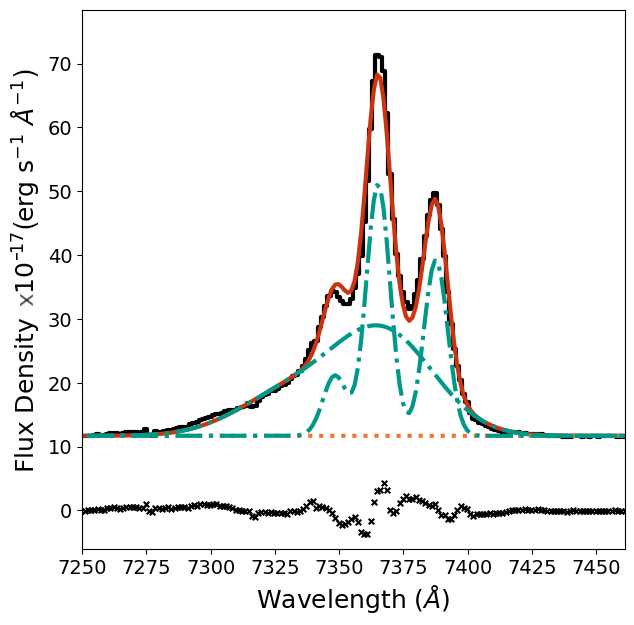}
        \vfill
        \includegraphics[width=0.435\textwidth]{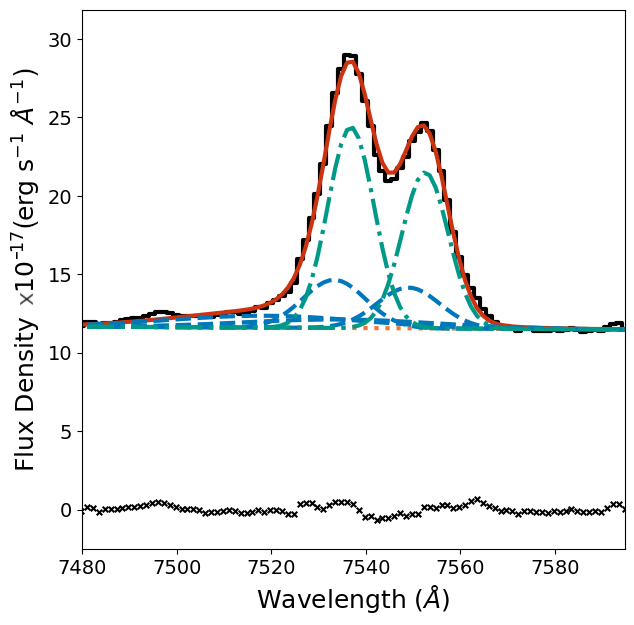}
        \hspace*{1.39cm}
        \includegraphics[width=0.435\textwidth]{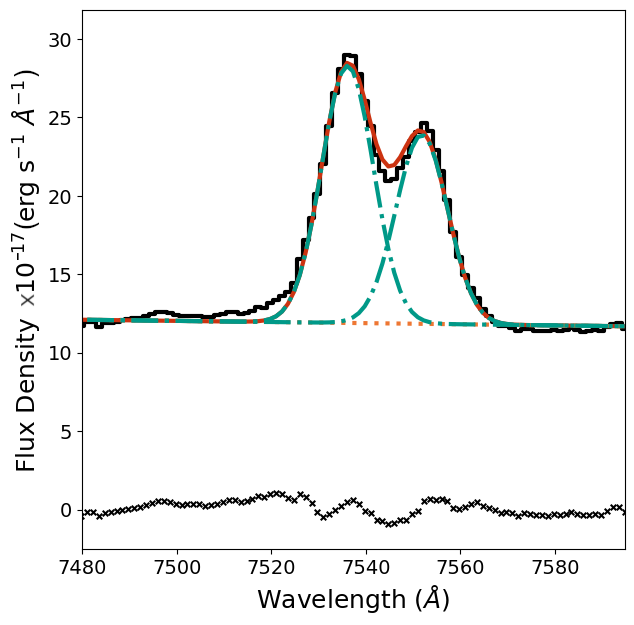}
    \end{subfigure}
    \caption{Fits to the [OIII]$\lambda\lambda4959,5007$ doublet (top rows), H$\alpha$ + {[}NII{]}$\lambda\lambda$6548,6584 blend (middle rows), and {[}SII{]}$\lambda\lambda$6717,6731 doublet (bottom rows) in Aperture 3 using the nuclear model + $N_\mathrm{g}$ Gaussian components (left column) and free-fitting (right column) approaches. The spectrum extracted from the aperture is shown as a black solid line, the overall fit in each case is shown as a solid red line, the first-order polynomial fit (accounting for the continuum) is shown as a dotted orange line, the components from the nuclear model (left panels only) are shown as a dashed blue line, and the additional Gaussian components (left panels) / free-fit components (right panels) are shown as green dash-dotted lines. The fits to these lines for the other apertures are presented in Appendix\;\ref{section: appendix: aperture_emission_line_fits}.}
    \label{fig: analysis_and_results: extended_emission: ap3_line_fits}
\end{figure*}

The emission-line fits produced by both methods (the nuclear model + $N_\mathrm{g}$ Gaussian components, and the free fits) for Aperture\;3 are shown in Figure\;\ref{fig: analysis_and_results: extended_emission: ap3_line_fits}; the fits for the other apertures are presented in Appendix \ref{section: appendix: aperture_emission_line_fits}. For each of the diagnostic lines ([OIII], [SII], H$\alpha$ + [NII]), the ratio of the posterior probabilities of the nuclear-model fits to those of the free fits was used to determine if the nuclear model was required: the more-complex model (i.e. the nuclear-model-fits, which consisted of the greater number of parameters) was selected if the ratio of its posterior probability to that of the less-complex model (the free fits) was greater than two. The results of this selection for each line blend in each aperture are given in Table\;\ref{tab: muse_f13451_1232: analysis_and_results: extended_emission: aperture_kinematics_energetics}. In the case where the nuclear model was preferred, only the additional (non-nuclear-component) Gaussian components were used to derive gas kinematics and velocities, while in the free-fitting case, all Gaussian components were used.

\subsubsection{Physical conditions and energetics of the extended emission}
\label{section: analysis_and_results: extended_emission: aperture_properties}

A major source of uncertainty in deriving warm-ionised mass outflow rates and kinetic powers is the electron density \citep{Rose2018, Revalski2021, Revalski2022, Holden2023, HoldenTadhunter2023, Speranza2024}: robust measurement requires diagnostics that are sensitive to a range of values and insensitive to emission-line blending effects, such as a technique pioneered by \citet{Holt2011} that makes use of the transauroral [OII]$\lambda\lambda7320,7331$ and [SII]$\lambda\lambda4069,4076$ lines. However, since not all of these lines are contained in the spectral coverage of MUSE, we instead determined the electron density for each aperture using the traditional [SII](6717/6731) ratio with strict measurement criteria, as was previously done in \citet{Holden2023} for the Seyfert\;2 galaxy IC\;5063: if a measured ratio was not 3$\sigma$ away from the theoretical lower or upper ratio limit (\mbox{0.44\;\textless\;[SII](6717/6731)\;\textless\;1.45}), then a $3\sigma$ limit was taken in the opposite direction (to give an upper or lower limit). In this way, we ensured that the measured densities were precise, and not subject to effects resulting from the ratio saturating at high or low values. The measured [SII] ratios (or upper/lower limits) were then used to determine electron densities using the \textsc{PyNeb Python} module. Due to the necessary lines (namely [OIII]$\lambda4363$) not being well detected in the apertures, it was not possible to measure electron temperatures for the extended gas. Therefore, a temperature of $T_e=10,000$\;K (typical for the warm ionised phase: \citealt{Holden2023, HoldenTadhunter2023}) was assumed in the electron density calculations.

Next, the flux of the H$\alpha$ line was determined by summing the flux of all Gaussian components that corresponded to the H$\alpha$ line in the fitted blends. Note that the H$\beta$ line itself was not measured due to its significantly lower flux compared to H$\alpha$, resulting in it not being well detected in the majority of the apertures. To estimate the H$\beta$ flux --- required to estimate gas masses --- the emissivity ratio $j_\mathrm{H\alpha}/j_\mathrm{H\beta}=2.863$ at $T_e=10,000$\;K and $n_e=10^2$\;cm$^{-3}$ \citep{Osterbrock2006} was used. Warm-ionised gas masses for each aperture were then calculated using the derived H$\beta$ luminosities with  
\begin{equation}
	M_\mathrm{out}=\frac{L(H\beta)m_p}{\alpha^\mathrm{eff}_\mathrm{H\beta}hv_\mathrm{H\beta}n_e},
\end{equation}
where $\alpha^\mathrm{eff}_\mathrm{H\beta}$ is the Case B recombination coefficient for H$\beta$ (taken to be $3.02\times10^{-14}$\;cm$^{-3}$\;s$^{-1}$ for $T_e=10,000$\;K and $n_e=10^2$\;cm$^{-3}$: \citealt{Osterbrock2006}), $m_p$ is the proton mass, and $n_e$ is the derived electron density for the aperture.

The gas kinematics in each aperture were determined by calculating the non-parametric percentile velocity shifts ($v_{10}$ and $v_{90}$) and widths ($\mathrm{W_{80}}$) of the [OIII]$\lambda\lambda4959,5007$ doublet for each aperture, which are given in Tables \ref{tab: muse_f13451_1232: analysis_and_results: extended_emission: aperture_kinematics_energetics} and \ref{tab: muse_f13451_1232: analysis_and_results: extended_emission: aperture_kinematics_energetics_free}. We also measured the velocity widths and shifts of the H$\alpha$ emission line in the apertures, which were found to be consistent (within $1\sigma$) with those derived from the [OIII] lines in all cases.

Under the assumption that the gas in the apertures is outflowing, the 10th-percentile velocity shift ($v_{10}$ or $v_{90}$) that had the highest value was taken to be the outflow velocity $v_\mathrm{out}$ --- this was done to ensure that the kinematics of any genuinely-outflowing gas would not be underestimated due to projection effects. These outflow velocities were subsequently used to calculate mass outflow rates with
\begin{equation}
	\dot{M}_\mathrm{out}=\frac{M_\mathrm{out}v_\mathrm{out}}{\Delta{R}},
\end{equation}
where ${\Delta}R$ is the radial extent of the given aperture in the direction from the nucleus. Kinetic powers were then determined using the estimated mass flow rates:
\begin{equation}
	\dot{E}=\frac{1}{2}{\dot{M}_\mathrm{out}v^2_\mathrm{out}}.
\end{equation}
Finally, outflow coupling efficiencies ($\epsilon_f$) for each aperture were estimated by calculating the ratio of the measured kinetic powers to the AGN bolometric luminosity of F13451+1232 ($L_\mathrm{bol}=4.8\times10^{45}$\;erg\;s$^{-1}$: \citealt{Rose2018}).

For the beam-smearing-corrected case (in which the nuclear model was used in the fits to the extended emission; Table\;\ref{tab: muse_f13451_1232: analysis_and_results: extended_emission: aperture_kinematics_energetics}), the kinematics (50\;\textless\;$|v_w|$\;\textless\;180\;km\;s$^{-1}$; 370\;\textless\;$|v_p|$\;\textless\;440\;km\;s$^{-1}$; 400\;\textless\;$\mathrm{W_{80}}$\;\textless\;470\;km\;s$^{-1}$) are modest, and the mass outflow rates, kinetic powers, and coupling efficiencies (0.14\;\textless\;$\dot{M}_\mathrm{out}$\;\textless\;$1.50$\;M$_\odot$\;yr$^{-1}$; $1.21\times10^{-4}$\;\textless\;$\epsilon_f$\;\textless\;$1.88\times10^{-3}$\;per\;cent) are far less than those of the compact nuclear outflows detected in the warm ionised ($\dot{M}_\mathrm{out, [OIII]}=11.3$\;M$_\odot$\;yr$^{-1}$; $\epsilon_{f,\mathrm{ [OIII]}}=0.5$\;per\;cent of $L_\mathrm{bol}$: \citealt{Rose2018}), neutral atomic ($\dot{M}_\mathrm{out, HI}\sim6$\;M$_\odot$\;yr$^{-1}$; $\epsilon_{f,\mathrm{ HI}}\sim0.04$\;per\;cent of $L_\mathrm{bol}$: \citealt{Morganti2013_4c1250, Holden2024}), and cold molecular ($\dot{M}_\mathrm{out, CO(1-0)}\sim230$\;M$_\odot$\;yr$^{-1}$; $\epsilon_{f,\mathrm{ CO(1-0)}}\sim1.4$\;per\;cent of $L_\mathrm{bol}$: \citealt{Holden2024}) phases\footnote{We note that the combined kinetic power for the multiphase nuclear outflows ($\dot{E}_\mathrm{kin}\sim2$\;per\;cent of $L_\mathrm{bol}$) is significantly greater than the observable kinetic power of $T<10^{4.5}$\;K outflowing gas ($\dot{E}_\mathrm{kin}\sim0.04$\;per\;cent of $L_\mathrm{bol}$) on $<$4\;kpc scales in the hydrodynamical simulations of \citet{Ward2024}.}. For all apertures, the inclusion of the nuclear model when fitting the extended emission produced statistically-better fits to the [OIII] emission lines, further demonstrating that beam-smeared nuclear-outflow emission contributes significant flux up to radial distances of $r\sim4$\;arcseconds ($r\sim9$\;kpc; i.e. the maximum radial distance of the apertures from the primary nucleus).

In contrast, when beam smearing is not accounted for --- demonstrated here by the free-fitting case (Table\;\ref{tab: muse_f13451_1232: analysis_and_results: extended_emission: aperture_kinematics_energetics_free}) --- the kinematics (40\;\textless\;$|v_w|$\;\textless\;740\;km\;s$^{-1}$; 490\;\textless\;$|v_p|$\;\textless\;2300\;km\;s$^{-1}$; 860\;\textless\;$\mathrm{W_{80}}$\;\textless\;2600\;km\;s$^{-1}$), mass outflow rates (up to $\dot{M}_\mathrm{out}=7.4\pm4.8$\;M$_\odot$\;yr$^{-1}$), and coupling efficiencies ($2.97\times10^{-4}<\epsilon_f<5.84\times10^{-2}$\;per\;cent) are significantly higher --- comparable to those of the nuclear warm-ionised outflow \citep{Rose2018}. However, even in this case, the mass outflow rates are still much lower than that of the total nuclear outflow in F13451+1232 ($\dot{M}_\mathrm{out}\sim250$\;M$_\odot$\;yr$^{-1}$ for the combined warm ionised, neutral atomic, and cold molecular phases: \citealt{Holden2024}).

Finally, we highlight that the mass outflow rates, kinetic powers, and coupling efficiencies presented here represent upper limits for any real, spatially-extended outflows that may be present in F13451+1232 because they were calculated under the assumption that the additional non-nuclear components represent outflowing gas, which (as argued in Section\;\ref{section: analysis_and_results: kinematics}) is likely not the case in reality.

\begin{table*}
    \begin{subtable}{\textwidth}
        \renewcommand{\arraystretch}{1.2}
        \centering
        \centerline{
        \begin{tabular}{cccccc}
		Aperture & Nuclear model required & $v_w$ (km\;s$^{-1}$) & $v_p$ (km\;s$^{-1}$) & $\mathrm{W_{80}}$ (km\;s$^{-1}$) & EW$_\mathrm{H\alpha}$ ({\AA}) \\
            \hline \\
            \multirow{3}{*}{1} & [OIII] &   \multirow{3}{*}{$88\pm2$} &   \multirow{3}{*}{$370\pm10$}   &   \multirow{3}{*}{$401\pm10$} & \multirow{3}{*}{$-19.5\pm1.2$}  \\
                &   --- &   &   &   \\
                &   --- &   &   &   \\
                &   &   &   &   \\
            \multirow{3}{*}{2} & [OIII] &   \multirow{3}{*}{$179\pm96$} &   \multirow{3}{*}{$437\pm174$}   &   \multirow{3}{*}{$401\pm175$} & \multirow{3}{*}{$-113.1\pm28.8$}  \\
                &   H$\alpha$ &   &   &   \\
                &   --- &   &   &   \\
                &   &   &   &   \\
            \multirow{3}{*}{3} & [OIII] &   \multirow{3}{*}{$141\pm1$} &   \multirow{3}{*}{$437\pm4$}   &   \multirow{3}{*}{$467\pm4$} & \multirow{3}{*}{$-40.8\pm3.7$}   \\
                &   --- &   &   &   \\
                &   [SII] &   &   &   \\
                &   &   &   &   \\
            \multirow{3}{*}{4} & [OIII] &   \multirow{3}{*}{$51\pm4$} &   \multirow{3}{*}{$370\pm17$}   &   \multirow{3}{*}{$467\pm27$} & \multirow{3}{*}{$-14.4\pm1.2$}   \\
                &   --- &   &   &   \\
                &   --- &   &   &   \\
                &   &   &   &   \\
            \multirow{3}{*}{5} & [OIII] &   \multirow{3}{*}{$73\pm3$} &   \multirow{3}{*}{$370\pm14$}   &   \multirow{3}{*}{$467\pm15$} & \multirow{3}{*}{$-11.1\pm0.9$}   \\
                &   H$\alpha$ &   &   &   \\
                &   --- &   &   &   \\
                &   &   &   &   \\
        \end{tabular}
        }
    \end{subtable} \\
    \centering
    \begin{subtable}{\textwidth}
        \renewcommand{\arraystretch}{1.2}
        \centering
        \begin{tabular}{cccccc}
            Aperture & log$_{10}$($n_e$[cm$^{-3}$]) & $M$ ($\times10^6$\;M$_\odot$) & $\dot{M}$ (M$_\odot$\;yr$^{-1}$) & $\dot{E}_\mathrm{kin}$ ($\times10^{40}$ erg\;s$^{-1}$) & $\epsilon_f$ ($\times10^{-4}$ per\;cent) \\
            \hline \\
            1   & $2.53^{+0.34}_{-0.64}$ & $1.57\pm0.42$ & $0.14\pm0.04$ & $0.58\pm0.16$ & $1.21\pm0.33$  \\
            2   & \textless\;2.53 & \textgreater\;$4.10$ & \textgreater\;0.94 & \textgreater\;$4.07$ & \textgreater\;$8.48$   \\
            3   & $2.24^{+0.36}_{-0.62}$ & $29.5\pm8.3$ & $1.50\pm0.42$ & $9.04\pm0.91$ & $18.8\pm3.5$   \\
            4   & \textless\;$2.41$ & \textgreater\;$2.34$ & \textgreater\;0.20 & \textgreater\;$0.87$ & \textgreater\;$1.81$ \\
            5   & \textless\;$2.42$ & \textgreater\;$1.74$ & \textgreater\;0.15 & \textgreater\;$0.65$ & \textgreater\;$1.35$ \\
        \end{tabular} \\
    \end{subtable} \\
    \caption{Beam-smearing-corrected [OIII] kinematics (flux weighted velocity shift, percentile velocity shift, and non-parametric velocity width $\mathrm{W_{80}}$), electron densities, masses, flow rates, kinetic powers, and coupling efficiencies for the apertures extracted from the MUSE-DEEP datacube (Figure\;\ref{fig: analysis_and_results: extended_emission: apertures}). The `Nuclear model required' column indicates if including the nuclear models (e.g. Figure\;\ref{fig: analysis_and_results: seeing: nuclear_aperture_spectrum}) in the modelling of the [OIII]$\lambda\lambda4959,5007$, H$\alpha$ + [NII], and [SII]$\lambda\lambda6717,6731$ line profiles produced better fits --- if so, only the additional (non-nuclear-model) Gaussian components were used to derive gas properties. Equivalent widths (EW) for the H$\alpha$ line are also given.}
    \label{tab: muse_f13451_1232: analysis_and_results: extended_emission: aperture_kinematics_energetics}
\end{table*}

\begin{table*}
    \begin{subtable}{\textwidth}
        \centering
        \renewcommand{\arraystretch}{1.2}
        \begin{tabular}{ccccc}
		Aperture & $v_w$ (km\;s$^{-1}$) & $v_p$ (km\;s$^{-1}$) & $\mathrm{W_{80}}$ (km\;s$^{-1}$) \\
            \hline \\
            1 & $-48\pm9$ & $-498\pm93$ & $868\pm116$ \\
            2 & $-109\pm3$ & $-898\pm19$ & $1335\pm21$ \\
            3 & $-266\pm4$ & $-1099\pm13$ & $1469\pm14$ \\
            4 & $-331\pm10$ & $-898\pm26$ & $1268\pm28$ \\
            5 & $-740\pm119$ & $-2300\pm249$ & $2600\pm250$ \\
        \end{tabular}
    \end{subtable} \\
    \vspace*{1cm} 
    \begin{subtable}{\textwidth}
        \centering
        \renewcommand{\arraystretch}{1.2}
        \begin{tabular}{cccccc}
            Aperture & log$_{10}$($n_e$[cm$^{-3}$]) & $M$ ($\times10^6$ M$_\odot$)  & $\dot{M}$ (M$_\odot$\;yr$^{-1}$) & $\dot{E}_\mathrm{kin}$ ($\times10^{40}$ erg\;s$^{-1}$)& $\epsilon_f$ ($\times10^{-4}$ per\;cent) \\
            \hline \\
            1   & $2.53^{+0.34}_{-0.64}$ & $1.57\pm0.42$ & $0.18\pm0.06$ & $1.42\pm0.54$ & $2.97\pm1.13$ \\
            2   & \textless\;2.53 & \textgreater\;4.28 & \textgreater\;2.99 & \textgreater\;76.1 & \textgreater\;159 \\
            3   & $1.95^{0.36}_{0.57}$ & $57.4\pm32.6$ & $7.36\pm4.79$ & $280\pm98$ & $584\pm240$ \\
            4   & \textless\;2.41 & \textgreater\;2.34 & \textgreater\;0.49 & \textgreater\;12.5 & \textgreater\;26.0 \\
            5   & \textless\;2.42 & \textgreater\;3.18 & \textgreater\;1.71 & \textgreater\;285 & \textgreater\;593 \\
        \end{tabular}
    \end{subtable} \\
    \caption{Non-beam-smearing-corrected [OIII] kinematics (flux weighted velocity shift, percentile velocity shift, and non-parametric velocity width $\mathrm{W_{80}}$), electron densities, and energetics (mass, flow rate, kinetic power, and coupling efficiencies) for the apertures extracted from the MUSE-DEEP datacube (Figure\;\ref{fig: analysis_and_results: extended_emission: apertures}). In this case, the nuclear models for [OIII]$\lambda\lambda4959,5007$, H$\alpha$ + [NII], and [SII]$\lambda\lambda6717,6731$ were not included in the fits to those lines.}
    \label{tab: muse_f13451_1232: analysis_and_results: extended_emission: aperture_kinematics_energetics_free}
\end{table*}

\subsubsection{The potential impact of underlying stellar continua and reddening}

Because the underlying stellar continuum in each aperture was not modelled and subtracted before emission-line fitting, the potential effect that this might have had on the resulting outflow properties was estimated by measuring the equivalent width of the Gaussian components of the H$\alpha$ line fits, which are given for each aperture in Table\;\ref{tab: muse_f13451_1232: analysis_and_results: extended_emission: aperture_kinematics_energetics}. Since the maximum expected EW for Balmer lines (in absorption) from modelling by \citet{GonzalezDelgado1999} is $\mathrm{EW}_\mathrm{H}\sim10$\;{\AA}, then in this case the impact on derived H$\alpha$ luminosity --- and hence outflow mass --- would be approximately a factor of two at most (for Aperture\;5, where the lowest EWs are measured), although we note that this is an upper limit. Importantly, when the H$\alpha$ luminosity for each aperture is corrected for an assumed stellar-absorption feature of $\mathrm{EW}_\mathrm{H}=10$\;{\AA}, the beam-smearing-corrected kinetic powers remain low ($\epsilon_f$\;\textless\;$2.5\times10^{-3}$\;per\;cent of $L_\mathrm{bol}$). Therefore, the lack of stellar-continuum modelling and correction does not affect the interpretations and conclusions made in this study.

In the majority of the apertures, the H$\beta$ emission line is not detected, and has low equivalent widths in those where it is. For this reason, we were unable to estimate the reddening using the Balmer decrement, which could be used to correct the line fluxes for extinction. \citet{Rose2018} estimated a reddening value of $\mathrm{E(B}-\mathrm{V)}\sim0.25$ for the `narrow' emission ($\mathrm{FWHM}\sim320$\;km\;s$^{-1}$) in the nucleus of F13451+1232 --- following the \citet{Cardelli1989} $R_\mathrm{v}=3.1$ extinction law, this corresponds to a flux-correction factor of $\sim$2 at the wavelength of H$\alpha$ (which we used to derive gas masses). Importantly, when applied to the values for the emission in the beam-smearing-corrected case, the maximum coupling efficiency is still orders of magnitude below the values used in models of galaxy evolution. Hence, correcting the measured line fluxes for reddening would not impact our interpretations and conclusions. Furthermore, we highlight that the extinction values at the locations of our selected apertures ($r>2$\;kpc) are likely much lower than that measured at the nucleus by \citet{Rose2018}, and therefore the real flux correction is likely much lower than a factor of $\sim2$. 

\section{Discussion}
\label{section: discussion}

By modelling the emission from compact outflows in the primary nucleus of F13451+1232 and accounting for this when fitting the spatially-extended [OIII]$\lambda\lambda4959,5007$, H$\alpha$ + {[}NII{]}$\lambda\lambda$6548,6584, and {[}SII{]}$\lambda\lambda$6717,6731 emission, it is found that the bulk of the warm-ionised gas has modest kinematics ($|v|$\;\textless\;300\;km\;s$^{-1}$; $\mathrm{W_{80}}$\;\textless\;500\;km\;s$^{-1}$: Figures\;\ref{fig: analysis_and_results: extended_emission: vw_maps} and \ref{fig: analysis_and_results: extended_emission: nuclear_subtracted_vw_maps}; Table\;\ref{tab: muse_f13451_1232: analysis_and_results: extended_emission: aperture_kinematics_energetics}) on large scales. We have argued that these kinematics are consistent with the gravitational motions that are expected in a major galaxy merger (Section\;\ref{section: analysis_and_results: kinematics}) and that therefore, this careful analysis of deep MUSE observations has found no evidence for galaxy-wide ($r$\;\textgreater\;5\;kpc) warm-ionised outflows in a type-2 quasar/ULIRG. Moreover, we have directly demonstrated that failure to account for the beam-smearing effects of atmospheric seeing would have lead to the incorrect interpretation of powerful, galaxy-wide outflows; if the large-scale gas were indeed outflowing, then correcting for beam smearing would significantly reduce measured outflow radii, kinematics, mass outflow rates, and kinetic powers. In this section, we compare our findings to those of previous observational studies, further highlight the importance of accounting for beam smearing when deriving the properties of AGN-driven outflows, and discuss the implications of the results presented here for models of galaxy evolution.

\subsection{Comparison to previous observational studies}
\label{section: discussion: comparison_to_observational_studies}


High-spatial resolution HST/ACS [OIII] imaging by \citet{Tadhunter2018} revealed compact ($r_\mathrm{[OIII]}\sim69$\;pc) warm-ionised emission near the primary nucleus of F13451+1232, consistent with outflow radii determined for the neutral atomic ($r_\mathrm{[HI]}$\;\textless\;100\;pc: \citealt{Morganti2013_4c1250}) and cold molecular ($r_\mathrm{CO(1-0)}$\;\textless\;120\;pc: \citealt{Holden2024}) phases. These results and the analysis of MUSE-DEEP data presented here thus provide evidence that, at least in these phases, the AGN-driven outflows in F13451+1232 are limited to a compact region around the primary nucleus, rather than being galaxy-wide. Given that the AGN in F13451+1232 may have recently restarted \citep{Stanghellini2005, Morganti2013_4c1250} and thus be young, it is possible that simply an insufficient amount of time has passed for the outflows to have reached large scales. In this context it is interesting to note that, in many active galaxies, the most kinematically-disturbed gas is located along the radio structure (e.g. \citealt{Ulvestad1981, Whittle1988, Morganti2007, Morganti2015, Venturi2021}) --- this is consistent with the scenario seen in F13451+1232, in which the compact radio structure and nuclear outflows exist on similar spatial scales ($r<100$\;pc: \citealt{Morganti2013_4c1250, Tadhunter2018, Holden2024}). \\

Deep IFU observations of other ULIRGs hosting AGN (such as those in the QUADROS sample: \citealt{Rose2018, Tadhunter2018, Spence2018, Tadhunter2019}), as well as compact radio sources of a range of ages (see \citealt{Kukreti2023}), will be important for establishing if the outflow extents determined for F13451+1232 are representative of the wider population of this important subclass of active galaxy. In addition, given the highly uncertain timescales of AGN activity, outflow acceleration, and potential impact on star formation, studies of outflows in galaxies at different times in the AGN lifecycle (e.g. \citealt{Santoro2020, Baron2022a, Baron2022b, Kukreti2023}) will continue to play an important role in determining the overall impact of outflows on host galaxies, which may be culmunative over multiple AGN episodes (see discussion in \citealt{Harrison2024}).


Moreover, our results support those of long-slit spectroscopy studies of other active galaxies of various types --- including quasars and Seyfert galaxies \citep{Das2006, VillarMartin2016, Fischer2018, Holden2023, HoldenTadhunter2023}, ULIRGs \citep{Spence2016, Rose2018, Spence2018}, and compact steep spectrum/gigahertz-peaked spectrum radio AGN \citep{Santoro2020} --- which found maximum warm-ionised-outflow radii in the range 60\;pc\,--\,6\;kpc (although the majority find radii towards the lower end of this range). Further evidence for compact outflows has been produced by high-spatial-resolution HST [OIII]-imaging (e.g. \citealt{Tadhunter2018}) and spectroscopic (e.g. \citealt{Das2005, Das2006, Fischer2018, Tadhunter2019, HoldenTadhunter2023}) studies, in addition to a technique that makes use of infrared spectral energy distribution fitting of dust emission \citep{Baron2019a}. Overall, these results indicate that AGN-driven outflows are limited to the central regions of active galaxies.

\subsection{The impact of beam-smearing on outflow spatial extents and kinematics}
\label{section: discussion: impact_on_extents_and_kinematics}

In contrast to the findings of this work and previous analyses that use a range of techniques to measure outflow extents, some ground-based IFU studies of active galaxies of various types have claimed evidence for galaxy-wide outflows in the warm ionised phase \citep{Liu2013, Liu2014, Harrison2014, McElroy2015}, with radial extents of $r$\;\textgreater\;10\;kpc being reported in some cases. We highlight that, in the velocity maps (e.g. $W_\mathrm{70}$, $\mathrm{W_{80}}$) presented in those studies, the extended ($r$\;\textgreater\;1\;arcsecond; $r$\;\textgreater\;5\;kpc) regions of high velocity width ($\mathrm{W_{80}}>500$\;km\;s$^{-1}$) often appear to be circular, with little variation in the measured velocity width (as noted by \citealt{Husemann2016} for the results of \citealt{Liu2013, Liu2014}). These regions --- which have previously been considered to represent genuine galaxy-wide outflow emission --- bear striking resemblance to the maps produced in this work for the case in which beam smearing is not accounted for (left panels of Figure\;\ref{fig: analysis_and_results: extended_emission: vw_maps}). Considering that this region is not present when the beam smearing of nuclear-outflow emission is accounted for (Figures\;\ref{fig: analysis_and_results: extended_emission: vw_maps} and \ref{fig: analysis_and_results: extended_emission: nuclear_subtracted_vw_maps}), we argue that in many (if not all) cases, the galaxy-wide high-velocity kinematics claimed in previous IFU studies were, in reality, due to atmospheric-seeing effects. In this context, it is important to note that, in the case presented here, the seeing smears compact-outflow emission over a radius ($r>3.5$\;arcseconds) that is at least eight times that of the HWHM of the seeing disk ($\mathrm{HWHM}_\star=0.40\pm0.10$\;arcseconds: Section\;\ref{section: observations_and_data_reduction: seeing}). Therefore, we argue that an outflow cannot be claimed to be genuinely spatially-resolved based solely on its measured radius being greater than the HWHM of the seeing disk. Moreover, we highlight that the impact of seeing on derived outflow radii will be greater at larger redshifts since the spatial scale (kpc/arcsecond) increases with redshift.

\citet{Hainline2014} and \citet{Husemann2016} investigated the impact of atmospheric seeing on the measured spatial extents of extended narrow line regions (ENLRs) of quasars in the redshift range 0.4\;\textless\;$z$\;\textless\;0.7 based on long-slit and IFU observations, respectively. Those studies showed that failure to account for beam smearing can lead to overpredictions of ENLR radii by up to a factor of two (although in many cases there was little impact on ENLR radii: \citealt{Husemann2016})\footnote{We note that ENLRs may not represent outflows, but rather have a major contribution from AGN-photoionised gas that is moving gravitationally; as demonstrated in this work, the impact on outflow radii is much more significant.}. In addition, in reanalysing the dataset presented by \citet{Liu2014} to account for beam smearing, \citet{Husemann2016} showed that the high [OIII] velocity widths ($\mathrm{W_{80}}$\;\textgreater\;1000\;km\;s$^{-1}$) claimed on $r$\;\textgreater\;1\;kpc scales were actually due to the beam smearing of compact outflows. Similarly, our study directly shows that when the beam smearing of AGN-driven-outflow emission is accounted for, spatially-extended ($r$\;\textgreater\;1\;kpc) high-velocity emission is not detected, and that most of the kinematics display only moderate velocity widths ($\mathrm{W_{80}}$\;\textless\;500\;km\;s$^{-1}$) and low velocity shifts ($v_w$\;\textless\;210\;km\;s$^{-1}$) on these scales (Figures\;\ref{fig: analysis_and_results: extended_emission: vw_maps} and \ref{fig: analysis_and_results: extended_emission: nuclear_subtracted_vw_maps}). Furthermore, here we have directly quantified the radial extent of this effect to be $r>3.5$\;arcseconds for outflowing gas, and we have directly demonstrated that, had beam smearing not been considered, outflow radii may have been overestimated by $r\sim7.4$\;kpc --- more than two orders of magnitude larger than the true outflow extent ($r\sim69$\;pc: \citealt{Tadhunter2018}) and far greater than the maximum factor of two found for ENLRs by \citet{Hainline2014} and \citet{Husemann2016}.

{It is interesting that when analysing IFU observations of a sample of nearby type-2 quasars that have similar bolometric luminosities to F13451+1232, \citet{Speranza2024} measured a warm-ionised-outflow radial extent of $r\sim12.6$\;kpc for one object (see also \citealt{Greene2012}), although the majority of the sample presented outflow radii consistent with what is found here. While we argue that many previous claims of galaxy-wide outflows were likely the result of beam smearing, the large-scale outflow reported by \citet{Speranza2024} shows a highly elongated and asymmetric structure relative to the nucleus which cannot be explained as being due to beam smearing, indicating that it is genuinely extended on these scales.

\subsection{The impact of beam-smearing on mass outflow rates and kinetic powers}
\label{section: discussion: impact_on_energetics}

By considering two cases --- one in which beam smearing is accounted for (Table\;\ref{tab: muse_f13451_1232: analysis_and_results: extended_emission: aperture_kinematics_energetics}) and one in which it is not (Table\;\ref{tab: muse_f13451_1232: analysis_and_results: extended_emission: aperture_kinematics_energetics_free}) --- this work shows that the beam smearing of nuclear outflow emission can artificially increase the derived flow rates of extended emission by up to an order of magnitude, and kinetic powers by one-to-two orders of magnitude. This is in close agreement with the findings of \citet{Husemann2016}, who found that accounting for the beam smearing of compact emission decreased the derived energetics of ENLR gas by these factors. Therefore, our study directly demonstrates that accounting for the effect of beam smearing on ground-based IFU studies of AGN-driven outflows is crucial: not only can atmospheric seeing artificially spread emission across the FOV (which may be interpreted as powerful, galaxy-wide outflows), but in the case of genuinely-extended emission, can increase the derived energetics significantly. 

Further investigation of this effect would involve applying the beam-smearing-correction technique used here to the re-analysis of the ground-based IFU observations that were used by previous studies that claim the presence of galaxy-wide outflows. Alternatively, techniques that less sensitive to (or completely unaffected by) the beam-smearing effects of atmospheric seeing could be used, such as ground-based spectroastrometry or high-resolution space-based IFU observations. Regarding the latter, some James Webb Space Telescope (JWST) studies based on early-release IFU observations that focused on spatially-resolving AGN-driven outflows have claimed a range of outflow radii (e.g. $3.6<r<10$\;kpc: \citealt{Veilleux2023, DEugenio2024, Liu2024}), although these studies assumed less conservative criteria for identifying outflowing gas than we do here.

\subsection{Comparison to models of galaxy evolution} 
\label{section: discussion: comparison_to_models}

Many semi-analytic (e.g. \citealt{Silk1998, Fabian1999, King2003, Zubovas2014}) and hydrodynamical (e.g. \citealt{DiMatteo2005, Curtis2016, Barai2018, Costa2018, Costa2022, Zubovas2023}) models of galaxy evolution that include AGN-driven outflows as a feedback mechanism predict them to extend to galaxy-wide scales and heat, entrain, and eject material that is needed for star formation (however, see also \citealt{Ward2024}). Despite being a physical representation of the situation considered in such models --- and therefore an ideal laboratory to verify this prediction --- here, no evidence of galaxy-wide outflows is seen in the warm ionised gas phase in the QSO2/ULIRG F13451+1232. If taken as an indication that there are no galaxy-wide outflows at all in this object, then the results presented in this work contradict the predictions of the models, potentially indicating that the outflows will not have a significant, direct effect on the evolution of the system.

Conversely, it is \textit{possible} that galaxy-wide outflows are present in F13451+1232, but the deep MUSE observations were not sensitive to them. For example, there may be a galaxy-wide warm-ionised component that has a surface brightness below the limit of the observations. However, we note that, in this case, the fluxes measured here could be considered upper limits; given these fluxes and the outflow masses and kinetic powers that are derived (Table\;\ref{tab: muse_f13451_1232: analysis_and_results: extended_emission: aperture_kinematics_energetics}), unphysically low electron densities for the warm-ionised gas ($n_e$\;\textless\;$10^{-4}$\;cm$^{-3}$) would be required to produce kinetic powers that are consistent with the coupling efficiency values used in models ($\dot{E}_\mathrm{kin}$\;\textgreater\;0.5\;per\;cent of $L_\mathrm{bol}$). Alternatively, some hydrodynamical models (e.g. \citealt{Costa2015, Costa2018, Curtis2016, Barai2018, Ward2024}) predict galaxy-wide outflows in a tenuous ($n\ll10^1$\;cm$^{-3}$), highly-ionised hot phase ($T$\;\textgreater\;10$^6$\;cm$^{-3}$), which MUSE observations are not sensitive to. A direct test of this scenario would require X-ray spectroscopy of low-surface-brightness emission, which could be enabled by future facilities such as LYNX.

\section{Conclusions}
\label{section: conclusions}

By accounting for the beam smearing of compact-outflow emission in the analysis of deep MUSE observations of the QSO2/ULIRG F13451+1232, the following has been found.

\begin{itemize}
	\item There is no evidence for fast ($\mathrm{W_{80}}>500$\;km\;s$^{-1}$) spatially-extended, galaxy-wide ($r$\;\textgreater\;5\;kpc) warm-ionised outflows in F13451+1232, despite it representing the situation considered by models of galaxy evolution that predict galaxy-wide outflows. Considering that outflows have previously been detected and characterised on compact scales ($r$\;\textless\;100\;pc) in this object, this is taken as confirmation that the high-velocity outflows are limited to the central regions of the galaxy, although we cannot rule out lower-velocity outflows that have similar kinematics to what is expected from gravitational motions in a galaxy merger.
		\item Even if we make the extreme assumption that the spatially-extended, kinematically-modest gas is outflowing, the mass outflow rates and kinetic powers are far below both those of the nuclear outflows and the predictions of theoretical models. This indicates that, even if this gas is truly outflowing, it is likely to have little impact on the evolution of the system.
    \item Failure to account for atmospheric seeing in ground-based IFU studies of AGN-driven outflows can lead to beam-smeared compact emission being interpreted as galaxy-wide outflows. Crucially, the contribution from such a beam-smeared component is significant far beyond the FWHM of the seeing disk (beyond radial distances of at least 3.5\;arcseconds, corresponding to $7.4$\;kpc for F13451+1232), which can lead to outflow radii being overestimated by two orders of magnitude. Therefore, considering emission that is detected on radial scales that are larger than the FWHM (or HWHM) of the seeing disk to be spatially-resolved is not a robust interpretation.
    \item When the beam-smearing effects of atmospheric seeing are accounted for, the derived velocity shifts and widths of any real outflows are significantly lower, mass outflow rates decrease by up to an order of magnitude, and outflow kinetic powers decrease by up to two orders of magnitude.
\end{itemize}

Overall, the analysis and results presented here present further evidence that AGN-driven outflows can be limited to the central kiloparsecs of galaxies, and demonstrate that accounting for beam-smearing in ground-based IFU studies of AGN-driven outflows is essential.

\section*{Acknowledgements}

We thank the anonymous referee for their feedback, which helped to improve the quality of this manuscript. LRH acknowledges a University of Sheffield publication scholarship. LRH and CNT acknowledge support from STFC. Based on observations collected at the European Southern Observatory under ESO programme 0103.B-0391 and data obtained from the ESO Science Archive Facility with DOI(s) under \url{https://doi.eso.org/10.18727/archive/42}. This research has made use of the NASA/IPAC Infrared Science Archive, which is funded by the National Aeronautics and Space Administration and operated by the California Institute of Technology. LRH thanks Stuart Littlefair for his helpful discussion in the development of the Bayesian emission-line-fitting routine used in this work, and James R. Mullaney and David J. Rosario for their helpful discussion that improved the quality of this study. LRH and CNT thank Cristina Ramos Almeida for her feedback which helped to improve the quality of the discussion section.

\section*{Data availability}

The data used in this work is available from the ESO Science Archive Facility (\url{http://archive.eso.org/cms.html}) with Run/Program ID 0103.B-0391 (PI: Arribas) and Archive ID ADP.2019-10-11T08:03:08.399.



\bibliographystyle{mnras}
\bibliography{./precise_outflow_diagnostics.bib} 




\appendix

\section{Emission-line fits to the apertures}
\label{section: appendix: aperture_emission_line_fits}

Here, we present the results of both emission-line-fitting approaches (nuclear model + $N_g$ gaussians, and free-fitting: see Section \ref{section: analysis_and_results: extended_emission: aperture_line_fits}) for the [OIII]$\lambda\lambda4959,5007$, [SII]$\lambda\lambda6717,6731$, and H$\alpha$+[NII]$\lambda\lambda6548,6584$ lines in each of the apertures extracted from the MUSE-DEEP datacube of F13451+1232 (Figure\;\ref{fig: analysis_and_results: extended_emission: apertures}); the fits for Aperture\;3 are presented in Figure\;\ref{fig: analysis_and_results: extended_emission: ap3_line_fits}. 

\begin{figure*}
    \centering
    \caption{Fits to the [OIII]$\lambda\lambda4959,5007$ doublet (top rows), H$\alpha$ + {[}NII{]}$\lambda\lambda$6548,6584 blend (middle rows), and {[}SII{]}$\lambda\lambda$6717,6731 doublet (bottom rows) in the apertures using the nuclear model + $N_\mathrm{g}$ Gaussian components (left column) and free-fitting (right column) approaches. The spectrum extracted from the aperture is shown as a black solid line, the overall fit in each case is shown as a solid red line, the first-order polynomial fit (accounting for the continuum) is shown as a dotted orange line, the components from the nuclear model (left panels only) are shown as a dashed blue line, and the additional Gaussian components (left panels) / free-fit components (right panels) are shown as green dash-dotted lines.\\}
    \label{fig: analysis_and_results: extended_emission: aperture_line_fits}
    \begin{subfigure}[t]{0.85\linewidth}
        \begin{minipage}{0.48\linewidth}
            \centering
            \textbf{Nuclear-model fits}
        \end{minipage}
        \hfill
        \begin{minipage}{0.42\linewidth}
            \centering
            \textbf{Free fits}
        \end{minipage}
        \vfill
        \includegraphics[width=0.45\textwidth]{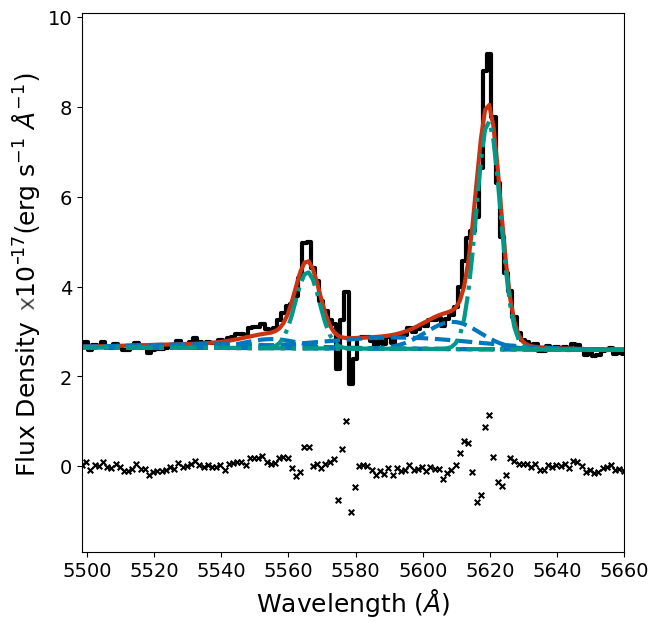}
        \hfill
        \includegraphics[width=0.45\textwidth]{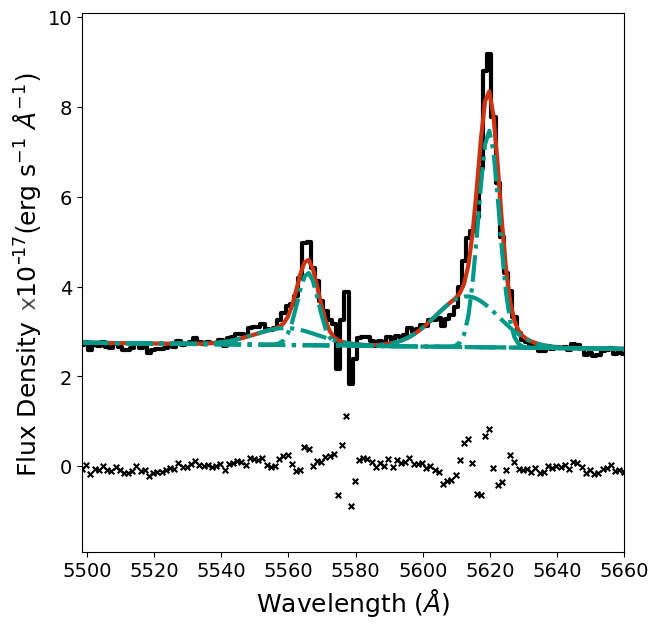}
        \vfill
        \includegraphics[width=0.435\textwidth]{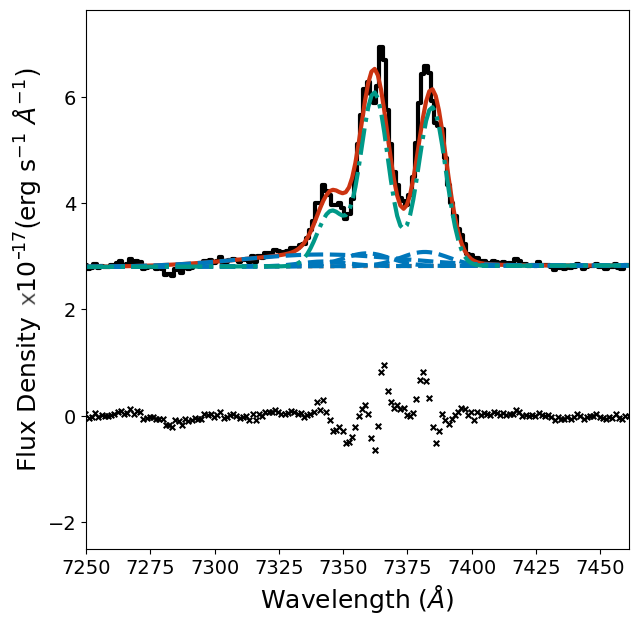}
        \hspace{1.42cm}
        \includegraphics[width=0.435\textwidth]{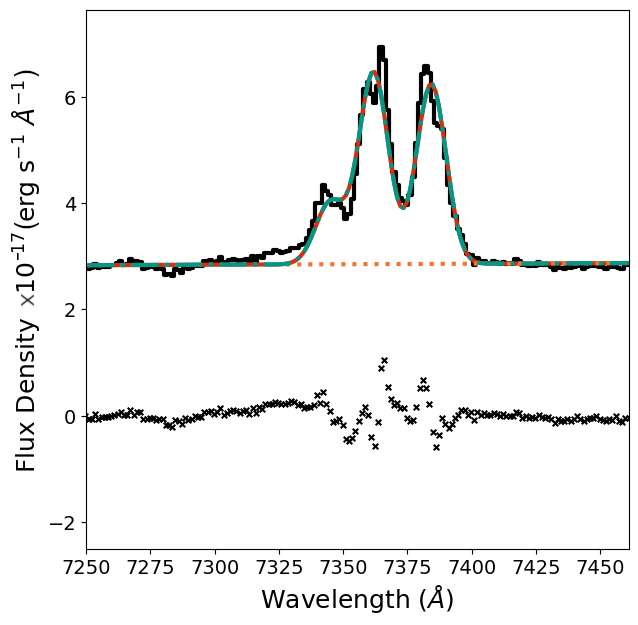}
        \vfill
        \includegraphics[width=0.45\textwidth]{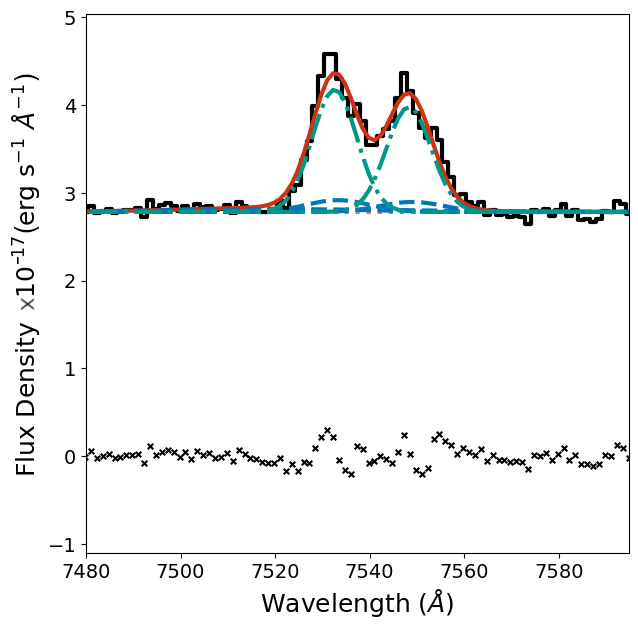}
        \hspace{1.3cm}
        \includegraphics[width=0.45\textwidth]{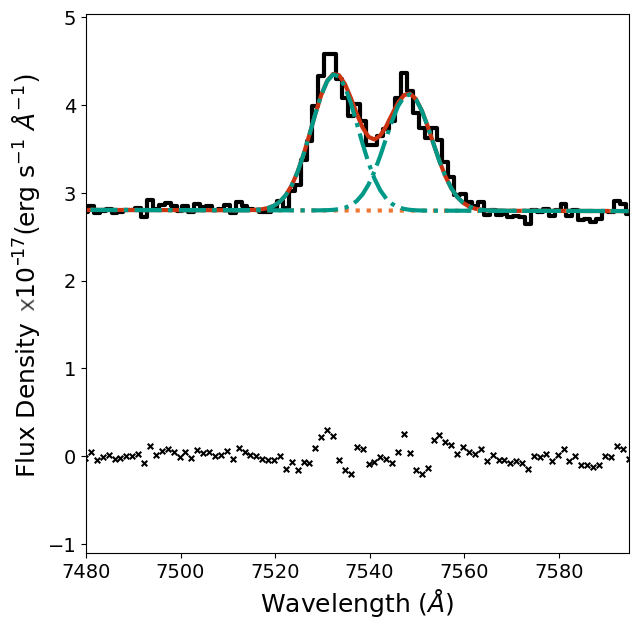}
        \label{fig: analysis_and_results: extended_emission: ap1_line_fits}
        \caption{The line fits for the spectrum extracted from Aperture 1. The feature at $\sim$5580\;{\AA} is instrumental, and did not have a significant impact on the resulting fits. }
    \end{subfigure}
\end{figure*}
\begin{figure*}\ContinuedFloat
    \centering
    \begin{subfigure}[t]{0.9\linewidth}
        \begin{minipage}{0.48\linewidth}
            \centering
            \textbf{Nuclear-model fits}
        \end{minipage}
        \hfill
        \begin{minipage}{0.42\linewidth}
            \centering
            \textbf{Free fits}
        \end{minipage}
        \vfill
        \includegraphics[width=0.45\textwidth]{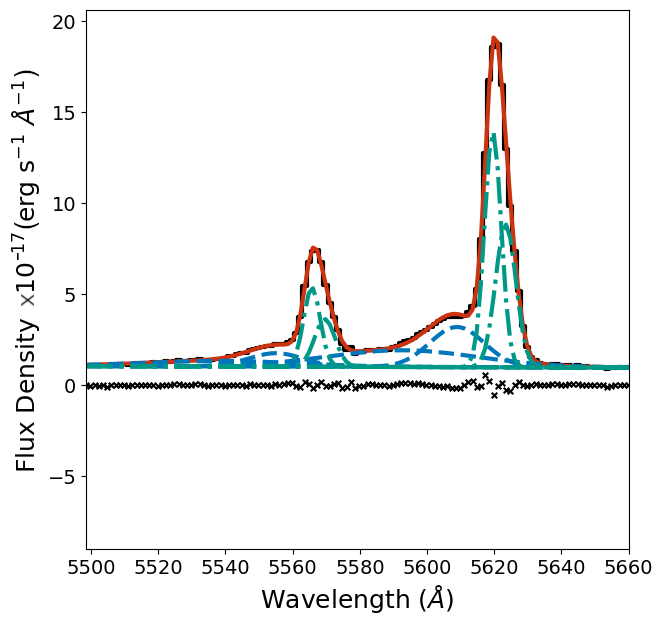}
        \hfill
        \includegraphics[width=0.45\textwidth]{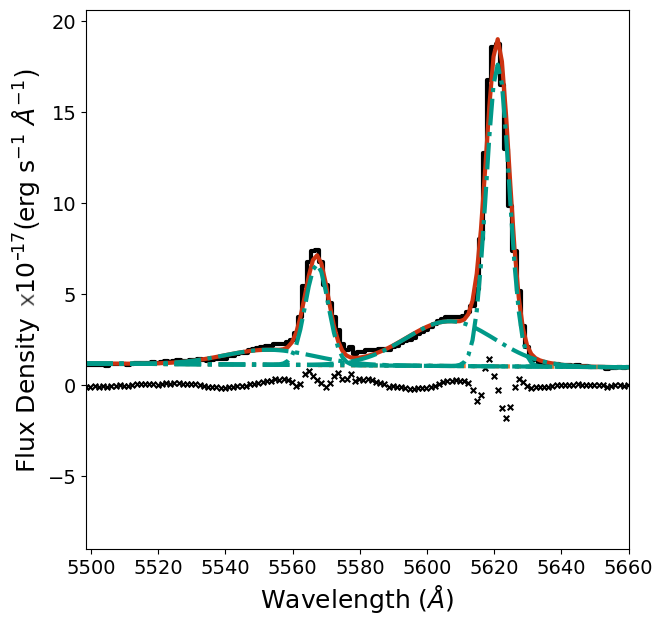}
        \vfill
        \includegraphics[width=0.435\textwidth]{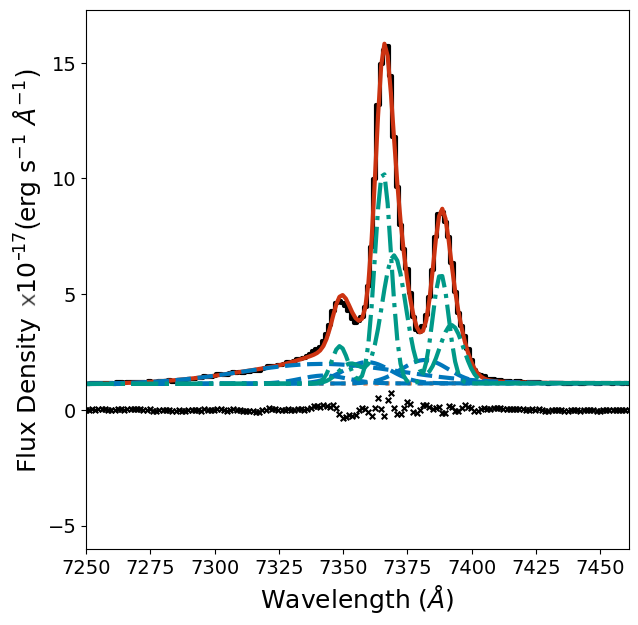}
        \hspace{1.42cm}
        \includegraphics[width=0.435\textwidth]{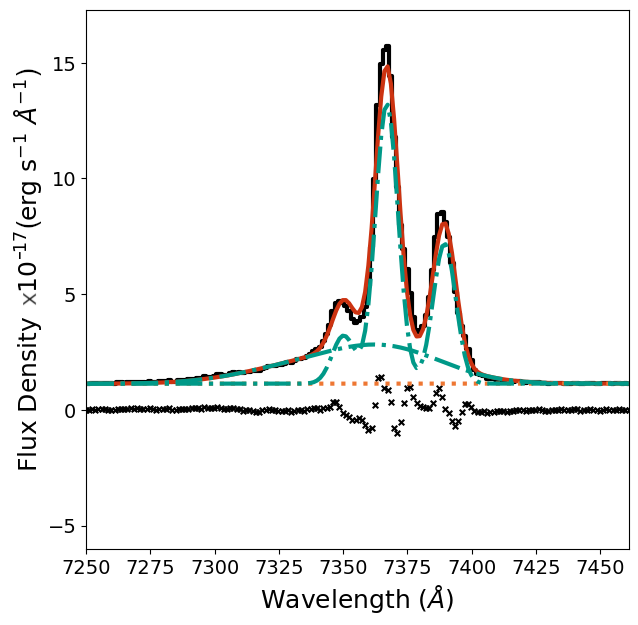}
        \vfill
        \includegraphics[width=0.45\textwidth]{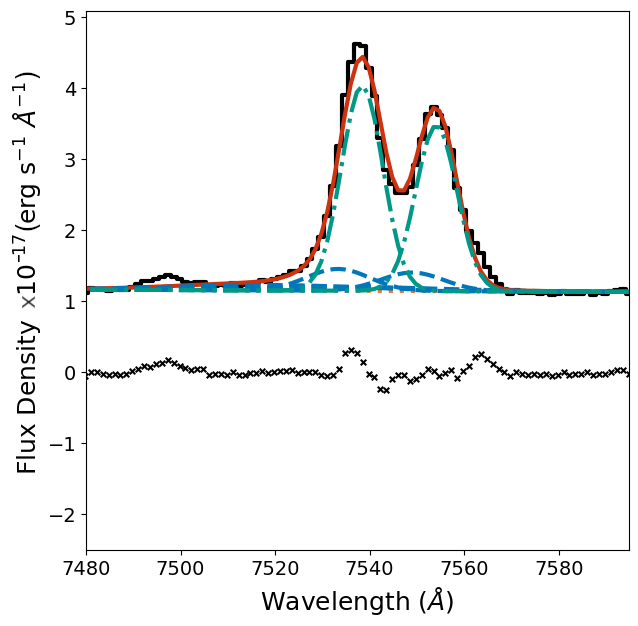}
        \hspace{1.3cm}
        \includegraphics[width=0.45\textwidth]{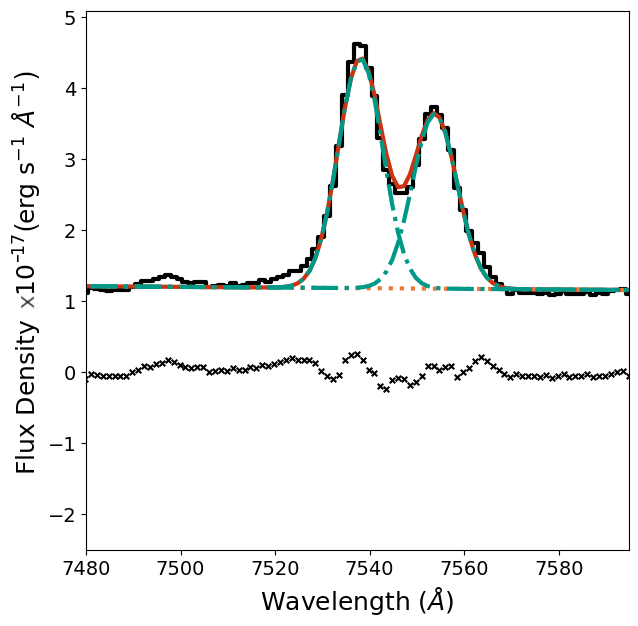}
        \label{fig: analysis_and_results: extended_emission: ap2_line_fits}
        \caption{The line fits for the spectrum extracted from Aperture 2.}
    \end{subfigure}
\end{figure*}
\begin{figure*}\ContinuedFloat
    \centering
    \begin{subfigure}[t]{0.9\linewidth}
        \begin{minipage}{0.48\linewidth}
            \centering
            \textbf{Nuclear-model fits}
        \end{minipage}
        \hfill
        \begin{minipage}{0.42\linewidth}
            \centering
            \textbf{Free fits}
        \end{minipage}
        \vfill
        \includegraphics[width=0.45\textwidth]{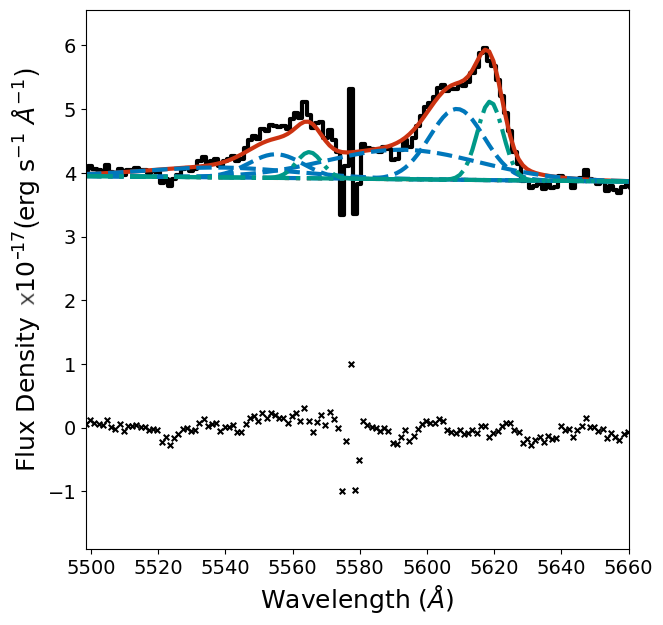}
        \hfill
        \includegraphics[width=0.45\textwidth]{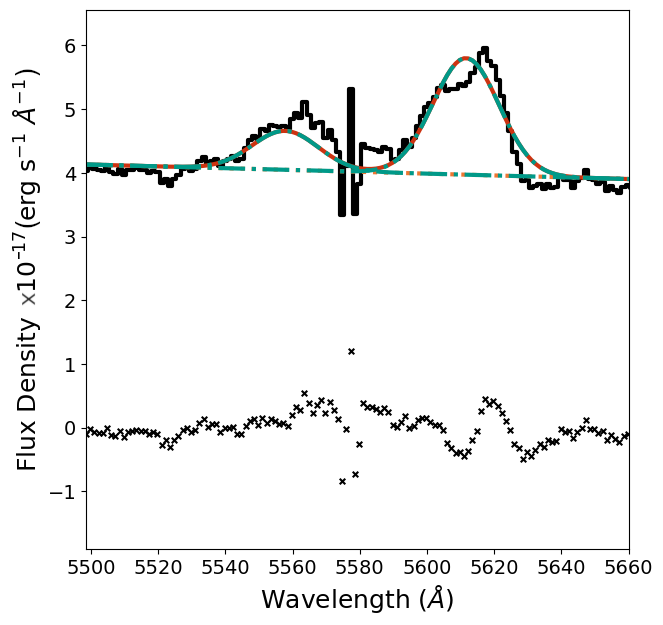}
        \vfill
        \includegraphics[width=0.435\textwidth]{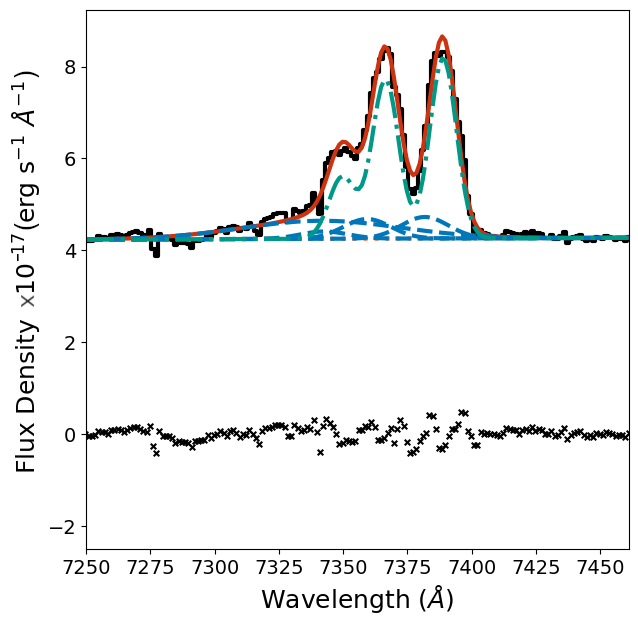}
        \hspace{1.42cm}
        \includegraphics[width=0.435\textwidth]{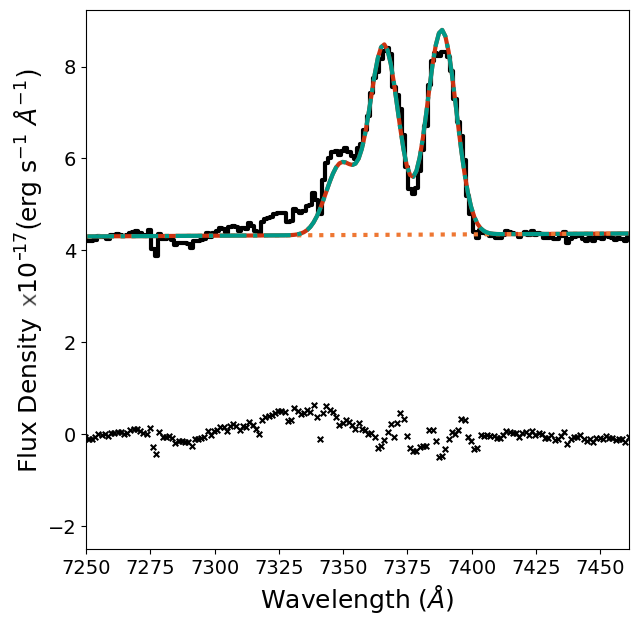}
        \vfill
        \includegraphics[width=0.45\textwidth]{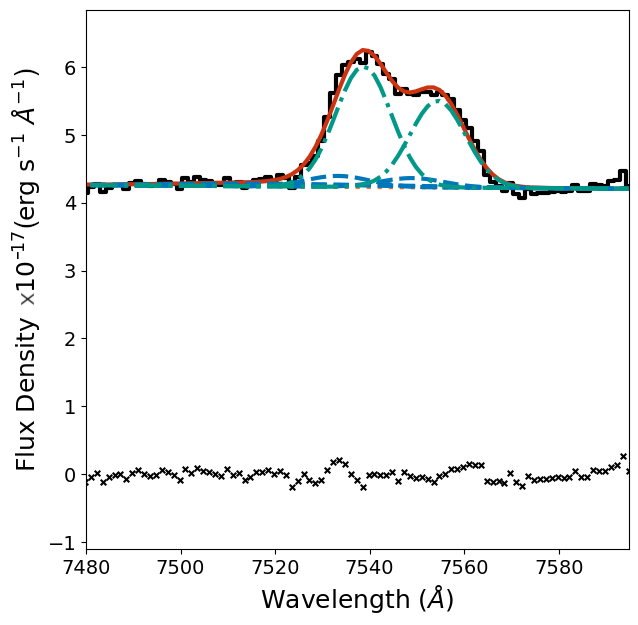}
        \hspace{1.3cm}
        \includegraphics[width=0.45\textwidth]{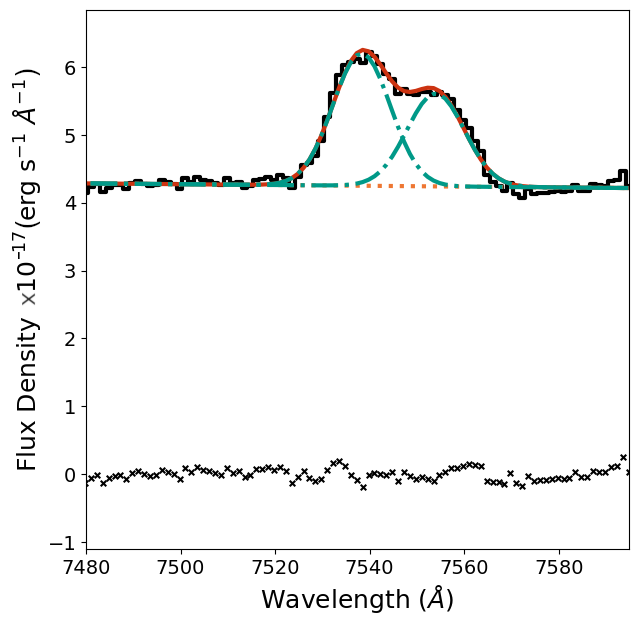}
        \label{fig: analysis_and_results: extended_emission: ap4_line_fits}
        \caption{The line fits for the spectrum extracted from Aperture 4.}
    \end{subfigure}
\end{figure*}
\begin{figure*}\ContinuedFloat
    \centering
    \begin{subfigure}[t]{0.9\linewidth}
        \begin{minipage}{0.48\linewidth}
            \centering
            \textbf{Nuclear-model fits}
        \end{minipage}
        \hfill
        \begin{minipage}{0.42\linewidth}
            \centering
            \textbf{Free fits}
        \end{minipage}
        \vfill
        \includegraphics[width=0.45\textwidth]{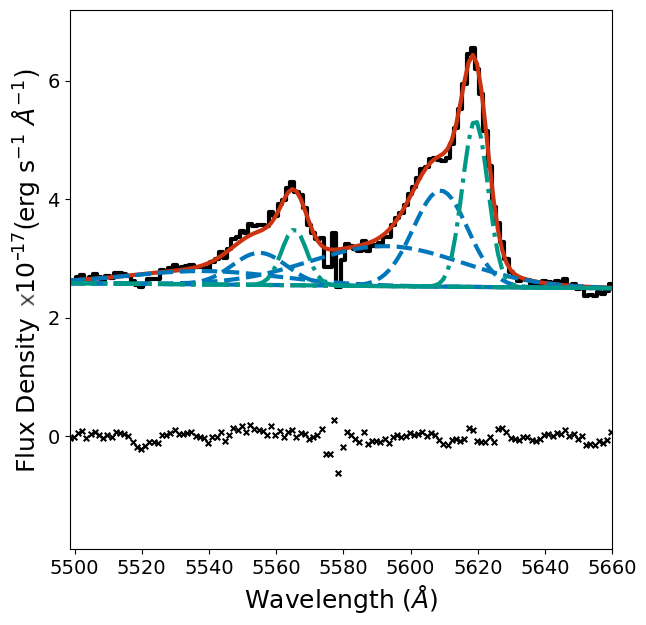}
        \hfill
        \includegraphics[width=0.45\textwidth]{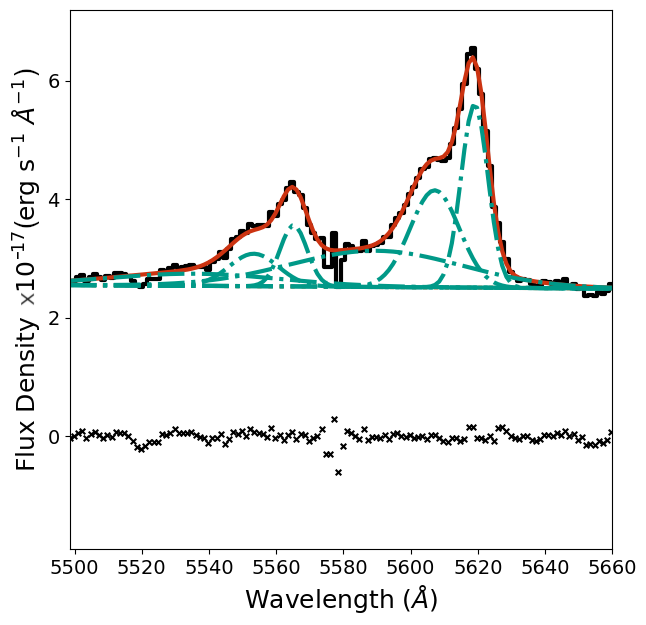}
        \vfill
        \includegraphics[width=0.435\textwidth]{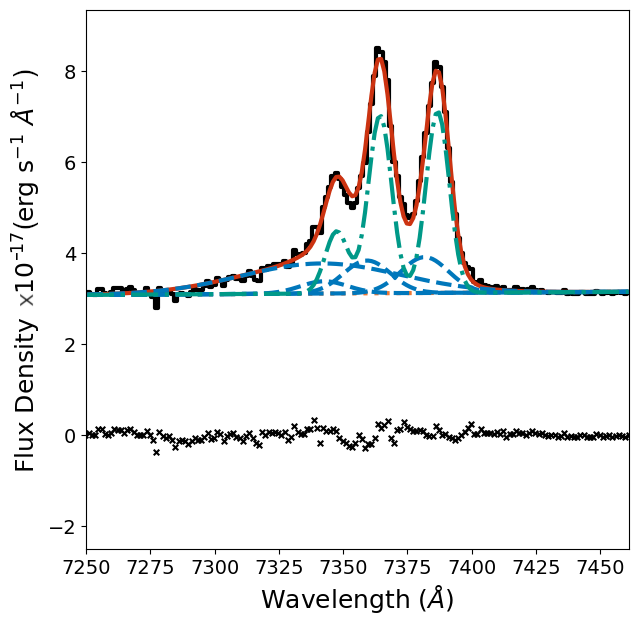}
        \hspace{1.42cm}
        \includegraphics[width=0.435\textwidth]{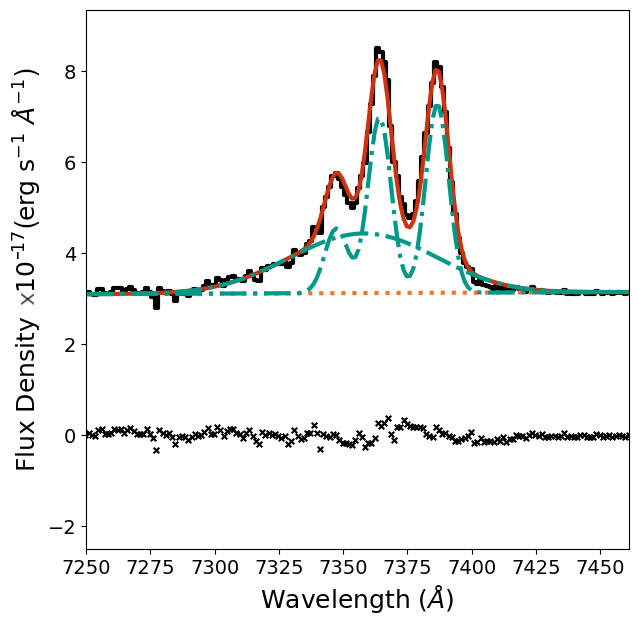}
        \vfill
        \includegraphics[width=0.45\textwidth]{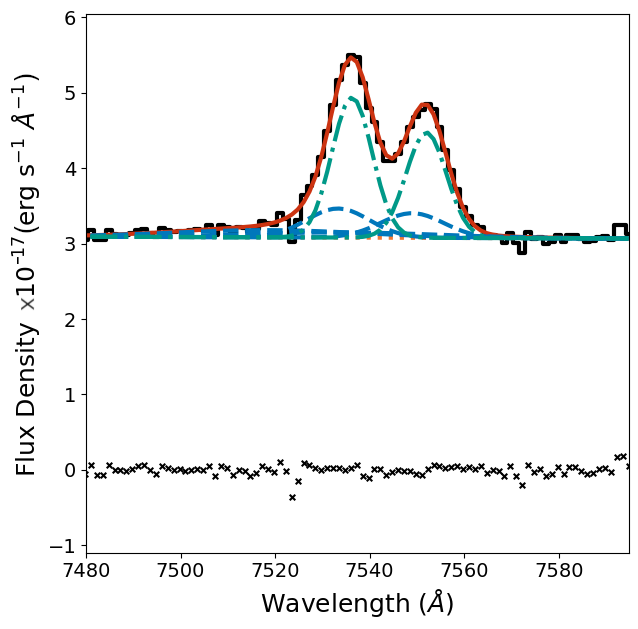}
        \hspace{1.3cm}
        \includegraphics[width=0.45\textwidth]{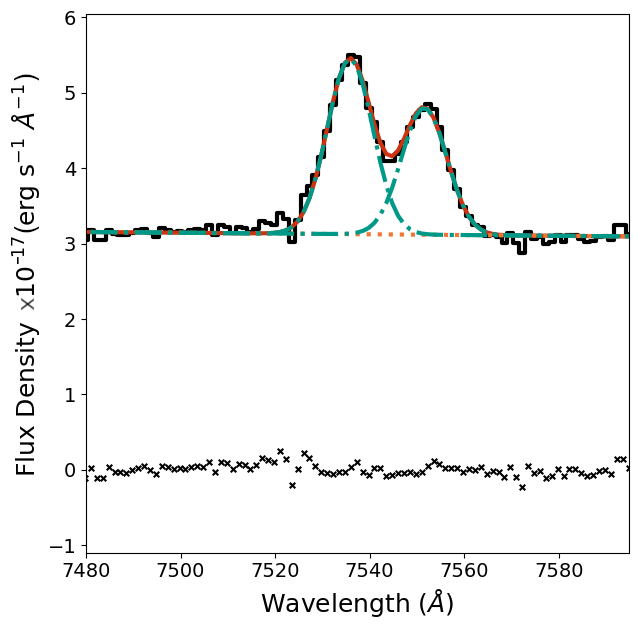}
        \label{fig: analysis_and_results: extended_emission: ap5_line_fits}
        \caption{The line fits for the spectrum extracted from Aperture 5.}
    \end{subfigure}
\end{figure*}


\bsp	
\label{lastpage}
\end{document}